\documentclass{elsart}
\usepackage{amsfonts}
\usepackage{amssymb}
\usepackage{amsmath}
\usepackage{graphicx}

\begin{document}

\begin{frontmatter}


\title{The isotropic XY model on the inhomogeneous periodic chain}


\author{J. P. de Lima$^{a,}$\thanksref{pim}},
\author{T. F. A. Alves$^{b}$},
\author{L. L. Gon\c{c}alves$^{a,}$\thanksref{lind}\corauthref{corresp}}
\address{$^{a}$Departamento de F\'{i}sica Geral, Instituto de F\'{i}sica, Universidade de S\~{a}o Paulo,
C.P.66318, 05315-970, S\~{a}o Paulo, SP, Brazil}
\address{$^{b}$Departamento de F\'{i}sica, Universidade Federal do Cear\'{a},
Campus do Pici, C.P.6030, 60451-970, Fortaleza, Cear\'{a}, Brazil}
\corauth[corresp]{Corresponding author: L. L. Gon\c{c}alves. Fax: +55-85-3288-9636\\
E-mail address:\emph{lindberg@fisica.ufc.br}}
\thanks[pim]{On a post-doctoral leave from Departamento de F\'{i}sica,
Universidade Federal do Piau\'{i}, Campus Ministro Petr\^{o}nio
Portela, 64049-550, Teresina, Piau\'{i}, Brazil.}
\thanks[lind]{On sabbatical leave from: Departamento de F\'{i}sica,
Universidade Federal do Cear\'{a}, Campus do Pici, C.P. 6030,
60451-970, Fortaleza, Cear\'{a}, Brazil.}

\begin{abstract}
The static and dynamic properties of the isotropic XY-model
$(s=1/2)$ on the inhomogeneous periodic chain, composed of \emph{N}
segments with \emph{n} different exchange interactions and magnetic
moments, in a transverse field \emph{h} are obtained exactly at
arbitrary temperatures. The properties are determined by introducing
the generalized Jordan-Wigner transformation and by reducing the
problem to a diagonalization of a finite matrix  of \emph{n-th}
order. The diagonalization procedure is discussed in detail and the
critical behaviour induced by the transverse field, at $T=0$, is
presented. The quantum transitions are determined by analyzing the
behaviour of the induced magnetization, defined as
$(1/n)\sum_{m=1}^{n}\mu_{m}<S_{j,m}^{z}>$ where $\mu_{m}$ is the
magnetic moment at site \emph{m} within the segment \emph{j}, as a
function of the field, and the critical fields determined exactly.
The dynamic correlations, $<S_{j,m}^{z}(t)S_{j^{'},m^{'}}^{z}(0)>$,
and the dynamic susceptibility $\chi_{q}^{zz}(\omega)$ are also
obtained at arbitrary temperatures. Explicit results are also
presented in the limit $T=0$, where the critical behaviour occurs,
for the static susceptibility $\chi_{q}^{zz}(0)$ as a function of
the transverse field \emph{h}, and for the frequency dependency of
dynamic susceptibility $\chi_{q}^{zz}(\omega)$. Also in this limit,
the transverse time-correlation
$<S_{j,m}^{x}(t)S_{j^{'},m^{'}}^{x}(0)>$, the dynamic
  and isothermal  susceptibilities, $\chi_{q}^{xx}(\omega)$ and $\chi_{T}^{xx}$, are obtained for the
transverse field greater or equal than the saturation field.

\end{abstract}

\end{frontmatter}

\section{Introduction}

Models involving inhomogeneous spin chains have been subject of
intensive study in recent years motivated by various reasons.
Amongst those, the most remarkable one is the necessity to
understand the unusual new properties
presented by low dimensional magnetic materials \cite%
{dagotto:1996,nguyen:1996,gambardella:2002,mukherjeea:2004} at low
temperature, which are described in terms of its many-body behaviour
and its quantum transitions \cite{sachdev:2000}. As manifestation of
these properties we can mention the appearance of magnetization
plateaus as functions of the external magnetic field
\cite{okamoto:2002,bostrem:2003}, the existence of an energy gap
between the ground state and first excited state at zero field
\cite{kramp:2000} and the presence of quantum critical behaviour
\cite{mukherjeea:2004,coleman:2001,matsumoto:2004}.

The one-dimensional XY model $(s=1/2)$ introduced by Lieb \textit{et
al}.\ \cite{lieb:1961} plays an important role in this context since
it constitutes one of the few many-body problems which can be
exactly solved. The most recent results on the inhomogeneous
anisotropic model have been obtained by Derzhko \textit{et
al}.\cite{derzhko:2004} and are restricted to the thermodynamic
properties. A good review of the known results is also presented in
his work.

For the isotropic model, the thermodynamic properties have also been
obtained by Derzhko\cite{derzhko:2000} (see references therein for
thermodynamic properties of the alternating chain), and the study of
the dynamics and of the quantum critical behaviour has been
restricted to the alternating chain
\cite{kontorovich:1968,perk:1980,derzkho:2000dyn},\ and for the
alternating superlattice\cite{delima:2002}.

In this paper we will also consider the isotropic XY model in a
transverse
field on the inhomogeneous periodic chain consisting of $N$ unit cells with $%
n$ sites, which has been studied by Derzkho \cite{derzhko:2000}.
Within the cells we can have $n$ different exchange constants as
well as magnetic moments, and the model corresponds to an extension
of the alternating superlattice. An extensive study of the quantum
critical behaviour will be presented and the dynamic properties in
the field direction determined for arbitrary temperature. Within a
new formalism, we have been able to solve the model exactly, and
present new and more general results from those presented in the
previous papers\cite{delima:2002,delima:1999}. In particular, we
have been able to obtain the dynamic correlation in the xy-plane, at
$T=0$, for transverse field greater than the saturation field.

In section 2 we discuss in detail the diagonalization of the model
and present the analytical results for the chains composed of cells
with two, three and four sites. The excitation spectrum is also
obtained, by a different method, in Appendix A, and it is shown
under which conditions the gap in the excitation spectrum is
suppressed.

The induced magnetization and the isothermal susceptibility $\chi
_{T}^{zz}$ are obtained in section 3 and we also determine the
critical fields associated with the quantum second order phase
transitions. The static and dynamic correlations in the field
direction are obtained in section 4 and, in the section 5, we
determine the longitudinal dynamic susceptibility $\chi
_{q}^{zz}(\omega )$. The dynamic correlation $\left\langle
S_{1,m}^{x}(t)S_{1+l,m^{\prime }}^{x}(0)\right\rangle $ is obtained,
at $T=0$ and for external fields greater than the saturation field,
in appendix B. Under these conditions, the isothermal and dynamic
transverse susceptibility $\chi ^{xx}$ are obtained in section 6,
and finally in section 7 we summarize the main results and present
the conclusions.

\section{The diagonalization of the Hamiltonian}

We consider the isotropic XY model $(s=1/2)$ on the inhomogeneous
periodic chain with $N$ cells, $n$ sites per cell, and lattice
parameter $a$, in a transverse field, whose unit cell is shown Fig.
1. The Hamiltonian is given
by%
\begin{eqnarray}
H &=&-\sum_{l=1}^{N}\left\{ \sum_{m=1}^{n}\mu
_{m}hS_{l,m}^{^{z}}+\sum_{m=1}^{n-1}J_{m}\left[
S_{l,m}^{x}S_{l,m+1}^{x}+S_{l,m}^{y}S_{l,m+1}^{y}\right] +\right.  \notag \\
&&\left.
+J_{n}S_{l,n}^{x}S_{l+1,1}^{x}+J_{n}S_{l,n}^{y}S_{l+1,1}^{y}\right\}
, \label{HamiltonianXX}
\end{eqnarray}%
where the parameters $J_{l,m}$ are the exchange coupling between
nearest-neighbour, $\mu _{m}$ the magnetic moments, $h$ the external
field and we have assumed periodic boundary conditions. If we
introduce the ladder
operators
\begin{figure}
\begin{center}
\includegraphics[width=14cm]{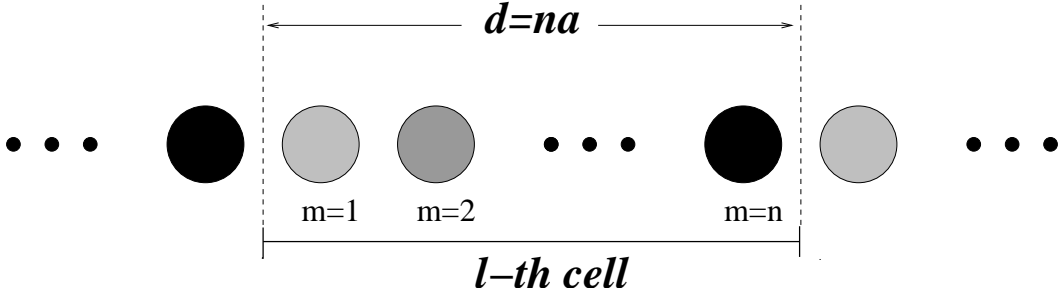}
\end{center}
\caption{Unit cell of the periodic chain.} \label{fig1}
\end{figure}
\begin{equation}
S^{\pm }=S^{x}\pm iS^{y},  \label{ladder}
\end{equation}%
and the generalized Jordan-Wigner
transformation\cite{goncalves:1995}
\begin{eqnarray}
S_{l,m}^{^{+}} &=&\exp \left\{ i\pi \sum_{l^{\prime
}=1}^{l-1}\sum_{m^{\prime }=1}^{n}c_{l^{\prime },m^{\prime
}}^{\dagger }c_{l^{\prime },m^{\prime }}+i\pi \sum_{m^{\prime
}=1}^{m-1}c_{l,m^{\prime
}}^{\dagger }c_{l,m^{\prime }}\right\} c_{l,m}^{\dagger },  \notag \\
S_{l,m}^{z} &=&c_{l,m}^{\dagger }c_{l,m}-\frac{1}{2},
\label{Jordan-Wigner}
\end{eqnarray}%
where $c_{l,m}$ and $c_{l,m}^{\dagger }$ are fermion annihilation
and creation operators, we can write the Hamiltonian
as\cite{siskens:1974}
\begin{equation}
H=H^{+}P^{+}+H^{-}P^{-},  \label{HamiltonianFermions}
\end{equation}%
where%
\begin{eqnarray}
H^{\pm } &=&-\sum_{l=1}^{N}\left\{ \sum_{m=1}^{n}\mu _{m}h\left(
c_{l,m}^{\dagger }c_{l,m}-\frac{1}{2}\right) +\sum_{m=1}^{n-1}\frac{J_{m}}{2}%
\left( c_{l,m}^{\dagger }c_{l,m+1}+c_{l,m+1}^{\dagger
}c_{l,m}\right)
\right\} -  \notag \\
&&-\sum_{l=1}^{N-1}\frac{J_{n}}{2}\left( c_{l,n}^{\dagger
}c_{l+1,1}+c_{l+1,1}^{\dagger }c_{l,n}\right) \pm
\frac{J_{n}}{2}\left( c_{N,n}^{\dagger }c_{1,1}+c_{1,1}^{\dagger
}c_{N,n}\right) ,
\end{eqnarray}%
and%
\begin{equation}
P^{\pm }=\frac{I\pm P}{2},
\end{equation}%
with $P$ given by
\begin{equation}
P=\exp \left( i\pi \sum_{l=1}^{N}\sum_{m=1}^{n}c_{l,m}^{\dagger
}c_{l,m}\right) .
\end{equation}%
As it is well known
\cite{siskens:1974,capel:1977,goncalvestese:1977}, since the
operator $P$ commutes with the Hamiltonian, the eigenstates have
definite parity, and $P^{-}(P^{+})$ corresponds to a projector into
a state of odd (even) parity.

Introducing periodic and anti-periodic boundary conditions on
$c^{\prime }s$ for $H^{-}$ and $H^{+}$ respectively, the
wave-vectors in the Fourier transform \cite{barbosafilho:2001},
\begin{eqnarray}
c_{l,m} &=&\frac{1}{\sqrt{N}}\sum_{q}\exp \left( -iqdl\right)
A_{q,m},
\notag \\
A_{q,m} &=&\frac{1}{\sqrt{N}}\sum_{l=1}^{N}\exp \left( iqdl\right)
c_{l,m}, \label{FourierXY}
\end{eqnarray}%
are given by $q^{-}=\frac{2r\pi }{Nd}$ for periodic condition and $q^{+}=%
\frac{\pi (2r+1)}{Nd}$, for anti-periodic condition, with $r=0,\pm
1,....,\pm N/2$, and $H^{-}$ and $H^{+}$ can be written in the form
\begin{equation}
H^{\pm }=\sum_{q^{\pm }}H_{q^{\pm }},  \label{Hamiltonianq}
\end{equation}%
where
\begin{eqnarray}
H_{q^{\pm }} &=&-\sum_{m=1}^{n}\mu _{m}h\left( A_{q^{\pm
},m}^{\dagger
}A_{q^{\pm },m}-\frac{1}{2}\right) -  \notag \\
&&-\sum_{m=1}^{n-1}\frac{J_{m}}{2}\left[ A_{q^{\pm },m}^{\dagger
}A_{q^{\pm
},m+1}+A_{q^{\pm },m+1}^{\dagger }A_{q^{\pm },m}\right] -  \notag \\
&&-\frac{J_{n}}{2}\left[ A_{q^{\pm },n}^{\dagger }A_{q^{\pm },1}\exp
(-idq^{\pm })+A_{q^{\pm },1}^{\dagger }A_{q^{\pm },n}\exp (idq^{\pm
})\right] .  \label{Hq}
\end{eqnarray}

Although $H^{-}$ and $H^{+}$ do not commute, it can be shown that in
the thermodynamic limit all the static properties of the system can
be obtained in terms of $H^{-}$ or $H^{+}.$ However, even in this
limit, some dynamic
properties depend on $H^{-}$ and $H^{+}$ \cite%
{siskens:1974,capel:1977,goncalvestese:1977}. Since%
\begin{equation}
\lbrack H_{q},H_{q^{\prime }}]=0,
\end{equation}%
where we make the identification $q\equiv q^{\pm },$ we can
diagonalize the
Hamiltonian by introducing the canonical transformation%
\begin{equation}
A_{q,m}=\sum_{k=1}^{n}u_{q,km}\xi _{q,k},\text{ \ \ \ \
}A_{q,m}^{\dag }=\sum_{k=1}^{n}u_{q,km}^{\ast }\xi _{q,k}^{\dag },
\label{transformationI}
\end{equation}%
and by imposing the condition
\begin{equation}
\lbrack \xi _{q,k},H_{q}]=\varepsilon _{q,k}\xi _{q,k},
\label{comutator}
\end{equation}%
which leads, for the coefficients $u_{q,km},$ to the equation
\begin{equation}
\mathbf{A}_{q}%
\begin{pmatrix}
u_{q,k1} \\
u_{q,k2} \\
\vdots \\
u_{q,kn}%
\end{pmatrix}%
=\varepsilon _{q,k}%
\begin{pmatrix}
u_{q,k1} \\
u_{q,k2} \\
\vdots \\
u_{q,kn}%
\end{pmatrix}%
,  \label{eigenequation}
\end{equation}%
where $\mathbf{A}_{q}$ is given by%
\begin{equation}
\mathbf{A}_{q}\equiv -%
\begin{pmatrix}
h_{1} & \frac{J_{1}}{2} & 0 & \cdots & 0 & \frac{J_{n}}{2}\exp
\left(
-iqd\right) \\
\frac{J_{1}}{2} & h_{2} & \frac{J_{2}}{2} &  &  & 0 \\
0 & \frac{J_{2}}{2} & h_{3} & \frac{J_{3}}{2} &  & \vdots \\
\vdots &  & \frac{J_{3}}{2} & \ddots & \ddots & 0 \\
0 &  &  & \ddots & h_{n-1} & \frac{J_{n-1}}{2} \\
\frac{J_{n}}{2}\exp \left( iqd\right) & 0 & \cdots & 0 &
\frac{J_{n-1}}{2} &
h_{n}%
\end{pmatrix}%
,  \label{Aq}
\end{equation}%
and the $u^{\prime }s$ satisfy the orthogonality relations

\begin{gather}
\sum\limits_{m=1}^{n}u_{q,km}u_{q,k^{\prime }m}^{\ast }=\delta
_{kk^{\prime
}},  \label{Ortogonality1} \\
\sum\limits_{k=1}^{n}u_{q,km}u_{q,km^{\prime }}^{\ast }=\delta
_{mm^{\prime }}.  \label{Ortogonality2}
\end{gather}%
Therefore the Hamiltonian can be written in the diagonal form
\begin{equation}
H_{q}=\sum_{k}\varepsilon _{q,k}(\xi _{q,k}^{\dagger }\xi _{q,k}-\frac{1}{2}%
),  \label{freefermions}
\end{equation}%
where the spectrum $\varepsilon _{q}$ of $H_{q}$ is determined from
the
determinantal equation%
\begin{equation}
det(\mathbf{A}_{q}-\varepsilon _{q}\mathbf{I})=0.
\label{dispersion}
\end{equation}%
In passing, we would like to note that for uniform magnetic moments,
$\mu _{m}\equiv \mu ,$ the term $-\sum_{l,m}^{n}\mu hS_{l,m}^{^{z}}$
commutes with the Hamiltonian, and consequently the effect of the
field is to shift the spectrum.

By using (\ref{Ortogonality1}) we can express the operators $\xi
^{\prime }s$ in terms of $A^{\prime }s$ which are given by
\begin{equation}
\xi _{q,k}=\sum\limits_{m=1}^{n}u_{q,km}^{\ast }A_{q,m}\text{
},\text{ \ \ \ \ \ \ \ \ \ }\xi _{q,k}^{\dagger
}=\sum\limits_{m=1}^{n}u_{q,km}A_{q,m\text{ }}^{\dagger },
\label{transformationinv}
\end{equation}%
and from eq.(\ref{FourierXY}) we obtain$,$
\begin{gather}
\xi
_{q,k}=\frac{1}{\sqrt{N}}\sum\limits_{l=1}^{N}\sum\limits_{m=1}^{n}\exp
\left( iqdl\right) u_{q,km}^{\ast }c_{l,m},  \notag \\
\xi _{q,k}^{\dagger }=\frac{1}{\sqrt{N}}\sum\limits_{l=1}^{N}\sum%
\limits_{m=1}^{n}\exp \left( -iqdl\right) u_{q,km}c_{l,m}^{\dagger
}, \label{directtransformation}
\end{gather}%
and their inverse, as
\begin{gather}
c_{l,m}=\frac{1}{\sqrt{N}}\sum\limits_{q}\sum\limits_{k}\exp \left(
-iqdl\right) u_{q,km}\xi _{q,k},  \notag \\
c_{l,m}^{\dagger
}=\frac{1}{\sqrt{N}}\sum\limits_{q}\sum\limits_{k}\exp \left(
iqdl\right) u_{q,km}^{\ast }\xi _{q,k}^{\dagger }.
\label{inversetransformation}
\end{gather}

The solution of eqs.(\ref{eigenequation}) and (\ref{dispersion}) can
be obtained analytically for $n\leq 4,$ and numerically for $n>4$.
In particular, the exact dispersion relations for the cases
$n=2,3,4$ are given below.

For $n=2$ is given by%
\begin{equation}
\varepsilon _{q}=-\frac{1}{2}(\mu _{1}+\mu _{2})h\pm
\frac{1}{2}\sqrt{(\mu _{1}-\mu
_{2})^{2}h^{2}+J_{1}^{2}+J_{2}^{2}+2J_{1}J_{2}\cos (2q)}\text{,}
\label{spectrum2}
\end{equation}%
and for $n=3$ by \cite{korn:1961}%
\begin{equation}
\varepsilon _{q}=-\frac{1}{3}(\mu _{1}+\mu _{2}+\mu
_{3})h+2\sqrt{-R}\cos \left( \frac{\theta _{q}+p\pi }{3}\right) ,
\label{spectrum3}
\end{equation}%
where $p=0,2,4$ and $\theta _{q}=\mathrm{arccos}\left( R_{q}/\sqrt{-R{}^{3}}%
\right) $, with $R$ and $R_{q}$ \ given by%
\begin{equation}
R=\frac{3a_{2}-a_{1}^{2}}{9},\text{ \ \ \ \ \ }R_{q}=\frac{%
9a_{2}a_{1}-27a_{q}-2a_{1}^{3}}{54},
\end{equation}%
where
\begin{eqnarray}
a_{1} &=&\left( \mu _{1}+\mu _{2}+\mu _{3}\right) h,  \notag \\
\text{ }a_{2} &=&\left( \mu _{1}\mu _{2}+\mu _{1}\mu _{3}+\mu
_{2}\mu _{3}\right) h^{2}-\frac{1}{4}\left(
J_{1}^{2}+J_{2}^{2}+J_{3}^{2}\right) ,
\notag \\
a_{q} &=&\mu _{1}\mu _{2}\mu _{3}h^{3}-\frac{1}{4}\left( \mu
_{1}J_{2}^{2}+\mu _{2}J_{3}^{2}+\mu _{3}J_{1}^{2}\right) h+\frac{1}{4}%
J_{1}J_{2}J_{3}\cos (3q).  \label{coef-3}
\end{eqnarray}%
For $n=4$, the four branches of the dispersion relation are obtained
from
the expression \cite{korn:1961}%
\begin{equation}
\varepsilon _{q}=-\frac{A}{4}+\frac{\mp \sqrt{A_{1}+2Y_{q}}\pm \sqrt{%
-(A_{1}+2Y_{q})-2(A_{1}\mp B_{1}/\sqrt{A_{1}+2Y_{q}})}}{2},\text{ }
\label{spectrum4}
\end{equation}%
by considering the sign combinations $(+++,$ $---,$ $-+-,$ $+-+),$ and where%
\begin{eqnarray}
A_{1} &=&-\frac{3A^{2}}{8}+B,\text{ \ \ \ }B_{1}=\frac{A^{3}}{8}-\frac{AB}{2}%
+C,\text{ \ \ \ }  \notag \\
Y_{q} &=&-\frac{5}{6}A_{1}-2\sqrt{\frac{-P_{q}}{3}}\cos
(\frac{\theta
_{q}+2\pi }{3}),\text{ \ \ \ \ \ }\theta _{q}=\arccos \left( \frac{S_{q}}{2}%
\sqrt{\frac{27}{-P_{q}^{3}}}\right) ,  \notag \\
P_{q} &=&-\frac{A_{1}{}^{2}}{12}-G_{q},\text{ \ \ \ \ \ \ \ \ \ \ }S_{q}=-%
\frac{A_{1}^{3}}{108}+\frac{A_{1}G_{q}}{3}-\frac{B_{1}^{2}}{8},  \notag \\
G_{q} &=&-\frac{3A^{4}}{256}+\frac{BA^{2}}{16}-\frac{AC}{4}+D_{q},
\end{eqnarray}%
and%
\begin{eqnarray}
A &=&(\mu _{1}+\mu _{2}+\mu _{3}+\mu _{4})h,  \notag \\
B &=&(\mu _{2}\mu _{1}+\mu _{4}\mu _{1}+\mu _{3}\mu _{1}+\mu _{4}\mu
_{2}+\mu _{3}\mu _{2}+\mu _{4}\mu _{3})h^{2}-  \notag \\
&&-\frac{(J_{1}{}^{2}+J_{2}{}^{2}+J_{3}{}^{2}+J_{4}{}^{2})}{4},  \notag \\
C &=&(\mu _{4}\mu _{3}\mu _{1}+\mu _{3}\mu _{2}\mu _{1}+\mu _{4}\mu
_{3}\mu
_{2}+\mu _{4}\mu _{2}\mu _{1})h^{3}-  \notag \\
&&-\frac{(J_{4}^{2}\mu _{3}+J_{2}^{2}\mu _{1}+J_{1}^{2}\mu
_{3}+J_{4}^{2}\mu _{2}+J_{2}^{2}\mu _{4}+J_{1}^{2}\mu
_{4}+J_{3}^{2}\mu _{1}+J_{3}^{2}\mu
_{2})h}{4},  \notag \\
D_{q} &=&\mu _{4}\mu _{3}\mu _{2}\mu _{1}h^{4}-\frac{(J_{4}^{2}\mu
_{3}\mu _{2}+J_{3}^{2}\mu _{2}\mu _{1}+J_{2}^{2}\mu _{4}\mu
_{1}+J_{1}^{2}\mu
_{4}\mu _{3})h^{2}}{4}+  \notag \\
&&+\frac{(J_{1}^{2}J_{3}^{2}+J_{4}^{2}J_{2}^{2})}{16}-\frac{%
J_{1}J_{2}J_{3}J_{4}}{8}\cos (4q).
\end{eqnarray}%
The excitation spectrum, given by eq.(\ref{dispersion}), can also be
obtained by\ a transfer matrix technique which leads to a different,
but equivalent, expression. This calculation is presented in
appendix \ A, and
it is expressed as%
\begin{equation}
\frac{\Im (\omega ,h)}{J_{1}J_{2}...J_{n}}\equiv trace[\mathbb{T}%
_{cell}(\omega ,h)]=2\cos (dq),  \label{spectrumsimetrico}
\end{equation}%
where $\mathbb{T}_{cell}(\omega ,h)$ is given by eq.(\ref{tcell}).
Although it is not shown in Appendix A, we have verified that $\Im
(\omega ,h)$
depends on the square of the exchange constants, $%
\{J_{1}^{2},J_{2}^{2},....,J_{n}^{2}\}.$ This means that the effect
of the change of the signs of the $J^{\prime }s$ is to introduce a
shift of $\pi /d$
in the spectrum wave-vectors, as can be seen in eq.(\ref{spectrumsimetrico}%
).\

As it is also shown in Appendix A, for zero external field, there is
no energy gap between the ground state and the first excited state,
for $n$ odd and arbitrary $J^{\prime }s.$ However, for $n$ even, the
gap in the spectrum
is only suppressed when the $J^{\prime }s$ satisfy the condition%
\begin{equation}
J_{1}J_{3}...J_{n-1}=J_{2}J_{4}...J_{n}.  \label{specialcondition}
\end{equation}%
This latter result can also be obtained from the criticality
condition for 1D\ random Ising model in a transverse field
\cite{pfeuty:1979} \ and the
equivalence between XY chain and two decoupled transverse field Ising chains%
\cite{jullien:1978,dfisher:1994}.

The excitation spectrum for a chain with $n=8,$ equal and different
magnetic moments, is shown in Fig. 2. As can be seen, when\ the
$J^{\prime }s$ satisfy eq.(\ref{specialcondition}) (continuous line)
there is no gap at zero field, whereas a gap is present when
eq.(\ref{specialcondition}) is not
satified (dot-dashed line). Although these results are presented for equal $%
\mu ^{\prime }s,$ they are still valid when the $\mu ^{\prime }s$
are different.

The dotted line represents the spectrum for nonzero field, different
$\mu
^{\prime }s$ and the $J^{\prime }s$ also satisfying eq.(\ref%
{specialcondition}). As can be seen, the spectrum shifts and opens a
gap at zero and boundary wave-vectors, which is a consequence of the
non-comutativity, in this case, of the field term with the
Hamiltonian.

\section{The induced magnetization and isothermal susceptibility $\protect%
\chi _{T}^{zz}$}

From eqs.(\ref{Jordan-Wigner}) and (\ref{inversetransformation}) we
can express the local induced magnetization $M_{l,m}^{z}$ as
\begin{equation}
M_{l,m}^{z}\equiv \mu _{m}\left\langle S_{l,m}^{z}\right\rangle
=\frac{\mu _{m}}{N}\sum\limits_{q,k}u_{q,km}^{\ast
}u_{q,km}n_{q,k}-\frac{\mu _{m}}{2}, \label{localSz}
\end{equation}%
where the occupation number $n_{q,k}$ is given by
\begin{equation}
n_{q,k}=\frac{1}{1+e^{\beta \varepsilon _{q,k}}},
\label{numberoccupation}
\end{equation}%
and the calculation can be done by considering $H=H^{-}$ \cite%
{siskens:1974,capel:1977,goncalvestese:1977}, since we are
interested in the thermodynamic limit.

We can define an average cell magnetization operator in the $z$ direction, $%
\tau _{l}^{z},$\ as
\begin{equation}
\tau _{l}^{z}\equiv \frac{1}{n}\sum_{m=1}^{n}\mu _{m}S_{l,m}^{z}=\frac{1}{n}%
\sum_{m=1}^{n}\mu _{m}\left( c_{l,m}^{\dagger
}c_{l,m}-\frac{1}{2}\right) , \label{tauzdefinition}
\end{equation}%
which corresponds to a generalization of the cell spin operator
defined in
the study of the alternating superlattice\cite{delima:1999}. By using eqs. (%
\ref{localSz}) and (\ref{tauzdefinition}) we can write the induced
magnetization per site as
\begin{equation}
M^{z}\equiv \left\langle \tau _{l}^{z}\right\rangle =\frac{1}{n}%
\sum_{m=1}^{n}\mu _{m}\left(
\frac{1}{N}\sum\limits_{q,k}u_{q,km}^{\ast
}u_{q,km}n_{q,k}-\frac{1}{2}\right) ,  \label{inducedmagnetization1}
\end{equation}%
which can be written in the form
\begin{equation}
M^{z}=-\frac{1}{2nN}\sum\limits_{q,k,m}\mu _{m}u_{q,km}^{\ast
}u_{q,km}\tanh \left( \frac{\beta \varepsilon _{q,k}}{2}\right) .
\label{inducedmagnetization2}
\end{equation}%
The isothermal susceptibility can be obtained from the expression
\begin{eqnarray}
\chi _{T}^{zz} &\equiv &\frac{1}{n}\frac{\partial M^{z}}{\partial h}=-\frac{1%
}{2nN}\left\{ \sum\limits_{q,k,m}\mu _{m}\frac{\partial u_{q,km}^{\ast }}{%
\partial h}u_{q,km}\tanh \left( \frac{\beta \varepsilon _{q,k}}{2}\right)
+\right.  \notag \\
&&\left. +\sum\limits_{q,k,m}\mu _{m}u_{q,km}^{\ast }\frac{\partial u_{q,km}%
}{\partial h}\tanh \left( \frac{\beta \varepsilon _{q,k}}{2}\right)
+\right.
\notag \\
&&\left. +\frac{\beta }{2}\sum\limits_{q,k,m}\mu _{m}u_{q,km}^{\ast }u_{q,km}%
\text{sech}^{2}\left( \frac{\beta \varepsilon _{q,k}}{2}\right) \frac{%
\partial \varepsilon _{q,k}}{\partial h}\right\} .
\label{susceptibilityisothermal}
\end{eqnarray}%
At $T=0,$ where the system presents quantum transitions, the induced
magnetization, obtained from eq. (\ref{inducedmagnetization2}) in the limit $%
T\rightarrow 0$, is given by
\begin{equation}
M^{z}=-\frac{1}{2nN}\sum\limits_{q,k,m}\mu _{m}u_{q,km}^{\ast }u_{q,km}%
\mathrm{sign}\left( \varepsilon _{q,k}\right) ,  \label{tauzmeanT0}
\end{equation}%
and from eq.(\ref{susceptibilityisothermal}) we obtain $\chi
_{T}^{zz},$ which diverges at the critical fields $h_{c}.$ We
identify the largest critical field as $h_{s},$ since for
$h\geqslant h_{s}$ the induced magnetization is saturated.

For identical magnetic moments, the average cell magnetization operator, $%
\tau _{l}^{z},$ is proportional to the average cell spin operator, $\frac{1}{%
n}\sum_{m=1}^{n}S_{l,m}^{z},$ and eqs (\ref{inducedmagnetization2}) and (\ref%
{tauzmeanT0}) can be written, respectively, as%
\begin{equation}
M^{z}=-\frac{1}{2nN}\sum\limits_{q,k,m}\mu \tanh \left( \frac{\beta
\varepsilon _{q,k}}{2}\right) ,
\label{inducedmagnetization1uniform}
\end{equation}%
\begin{equation}
M^{z}=-\frac{1}{2nN}\sum\limits_{q,k,m}\mu \mathrm{sign}\left(
\varepsilon _{q,k}\right) ,  \label{inducedmagnetization2uniform}
\end{equation}%
where $\mu _{m}=\mu $ for any $m.$

The results for $M^{z}$ and $\chi _{T}^{zz},$ at $T=0,$ are
presented in
Fig. 3 as functions of the field $h,$ for a chain with $n=8$ and identical $%
\mu ^{\prime }s.$ The continuous line and the dashed line correspond
to the cases where the exchange constants satisfy and do not satisfy
the condition shown in eq.(\ref{specialcondition}), respectively. As
we have shown for the alternating superlattice \cite{delima:1999},
the magnetization also presents plateaus which are limited by
critical fields $h_{c},$ where the isothermal susceptibility
diverges, which correspond to quantum phase transitions induced by
the field. The regions of plateaus, which we associate with
disordered regions \cite{delima:1999}, correspond to the gaps in the
excitation spectrum, and the critical fields are associated with the
zero-energy mode with the wave-vectors $q=0$ and $\pi /d.$

As it can also be seen, when there is no gap in the excitation
spectrum at zero field (continuous line), the zero magnetization
plateau, which is present when there is a gap (dashed line), is
suppressed, and consequently the total number of transitions induced
by the field is $n-1.$ It should be noted that the local
magnetization also presents plateaus and non-analytic behaviour at
the critical fields.

For different magnetic moments, the critical fields are obtained as
in the
previous case, and the results for $M^{z}$ and $\chi _{T}^{zz},$ also at $%
T=0,$ are shown in Fig. 4. The main difference from the previous
case is the fact that the magnetization does not present plateaus,
and the susceptibility $\chi _{T}^{zz}$ is different from zero at
both sides of the transition, although it diverges at one side only.
The local magnetization presents similar behaviour as the total
magnetization, and there is no suppression of the transition at zero
field when there is no gap in the excitation spectrum (continuous
line). The regions between two critical fields where the
susceptibility $\chi _{T}^{zz}$ is finite correspond to the gaps in
the spectrum. As in the case of equal magnetic moments, we
associated these regions with disordered regions.

It should also be noted that for $T>0$, all these transitions are
suppressed by the thermal fluctuations.

The exact expressions for the critical fields for $n=2,3$ and $4$,
when we have different magnetic moments, can be obtained from\ the
analytic expressions for the spectra given in
eqs.(\ref{spectrum2}),(\ref{spectrum3}) and (\ref{spectrum4}), by
considering $\varepsilon _{q}=0$ and $q=0$ or $\pi /d.$ \ The
explicit results are
\begin{equation}
\text{\ }h_{c1}=\frac{\left\vert J_{1}+J_{2}\right\vert }{2\sqrt{\mu
_{1}\mu
_{2}}},\text{ \ \ \ \ \ \ }h_{c2}=\frac{\left\vert J_{1}-J_{2}\right\vert }{2%
\sqrt{\mu _{1}\mu _{2}}}\text{, \ \ for }n=2\text{
\cite{kontorovich:1968}}, \label{criticalfields2}
\end{equation}%
\begin{eqnarray}
h_{c1} &=&2\sqrt{R_{1}^{\prime }}\cos \left( \frac{\theta ^{\prime }}{3}%
\right) ,\text{ \ }h_{c2}=2\sqrt{R_{1}^{\prime }}\left\vert \cos
\left(
\frac{\theta ^{\prime }+2\pi }{3}\right) \right\vert ,\text{ \ }  \notag \\
h_{c3} &=&2\sqrt{R_{1}^{\prime }}\left\vert \cos \left( \frac{\theta
^{\prime }+\pi }{3}\right) \right\vert ,\text{ for }n=3,
\label{criticalfields3}
\end{eqnarray}%
where
\begin{eqnarray}
\theta ^{\prime } &=&\mathrm{arccos}\left( \frac{R_{2}^{\prime }}{\sqrt{%
R_{1}^{\prime 3}}}\right) \text{ with }0\leq \theta ^{\prime }/3\leq \pi /2,%
\text{ \ \ }  \notag \\
\text{\ \ }R_{1}^{\prime } &=&\frac{J_{1}^{2}\mu _{3}+J_{2}^{2}\mu
_{1}+J_{3}^{2}\mu _{2}}{12\mu _{1}\mu _{2}\mu _{3}}\text{, }R_{2}^{\prime }=%
\frac{J_{1}J_{2}J_{3}}{8\mu _{1}\mu _{2}\mu _{3}},
\end{eqnarray}%
and%
\begin{equation}
h_{c1}=h_{c}^{(+\text{ }-)},\text{ \ }h_{c2}=h_{c}^{(+\text{ }+)},\text{ \ }%
h_{c3}=h_{c}^{(-\text{ }+)},\text{ \ }h_{c4}=h_{c}^{(-\text{ }-)}\text{ for }%
n=4,  \label{criticalfields4}
\end{equation}%
where%
\begin{eqnarray}
h_{c}^{(\pm \text{ }\pm )} &=&\left\{ \frac{1}{8}\frac{\mu _{1}\mu
_{2}J_{3}^{2}+\mu _{1}\mu _{4}J_{2}^{2}+\mu _{2}\mu
_{3}J_{4}^{2}+\mu _{3}\mu _{4}J_{1}^{2}}{\mu _{1}\mu _{2}\mu _{3}\mu
_{4}}\pm \right.  \notag
\\
&&\pm \frac{1}{2\mu _{1}\mu _{2}\mu _{3}\mu _{4}}\left\{
\frac{1}{16}(\mu _{1}\mu _{2}J_{3}^{2}+\mu _{1}\mu _{4}J_{2}^{2}+\mu
_{2}\mu
_{3}J_{4}^{2}+\mu _{3}\mu _{4}J_{1}^{2})^{2}-\right.  \notag \\
&&\left. \left. -4\mu _{1}\mu _{2}\mu _{3}\mu _{4}\left[
\frac{1}{16}\left(
J_{1}^{2}J_{3}^{2}+J_{2}^{2}J_{4}^{2}\right) \pm \frac{1}{8}%
J_{1}J_{2}J_{3}J_{4}\right] \right\} ^{1/2}\right\} ^{1/2}.
\end{eqnarray}

For arbitrary $n,$ the critical fields can also be obtained from the
solution of the equation%
\begin{equation}
\frac{\Im (0,h)}{2J_{1}J_{2}...J_{n}}\pm 2=0.
\end{equation}%
For homogeneous magnetic moments we can also obtain from eq. (\ref%
{tauzmeanT0}), at $T=0,$ analytical expressions for the induced
magnetization $M^{z}.$ For positive values of $J^{\prime }s,$ the
critical fields satisfy the relation $h_{ci}>h_{ci+1}$ and the
induced magnetization, for $n=2,$ is given explicitly by
\begin{equation}
M^{z}=\left\{
\begin{array}{lll}
\frac{\mu }{2} &  & h\geq h_{c1}\equiv h_{s}, \\
\frac{\mu }{2}-\frac{\mu }{2\pi }\arccos \Lambda _{2} &  &
h_{c2}\leq h\leq
h_{c1}, \\
0 &  & h\leq h_{c2},%
\end{array}%
\right.  \label{tauz2}
\end{equation}%
where
\begin{equation}
\Lambda _{2}=\frac{2\mu ^{2}h^{2}-\frac{1}{2}\left(
J_{1}^{2}+J_{2}^{2}\right) }{J_{1}J_{2}},  \label{lambda2}
\end{equation}%
with $h_{c1}$ and $h_{c2}$ given in eq.(\ref{criticalfields2}), and
for $n=3$
by%
\begin{equation}
M^{z}=\left\{
\begin{array}{lll}
\frac{\mu }{2} &  & h\geq h_{c1}\equiv h_{s}, \\
\frac{\mu }{2}-\frac{\mu }{3\pi }\arccos (-\Lambda _{3}) &  &
h_{c2}\leq
h\leq h_{c1}, \\
\frac{\mu }{6} &  & h_{c3}\leq h\leq h_{c2}, \\
\frac{\mu }{6}-\frac{\mu }{3\pi }\arccos \Lambda _{3} &  & h\leq h_{c3},%
\end{array}%
\right. \text{ }  \label{tauz3}
\end{equation}%
where
\begin{equation}
\Lambda _{3}=\frac{-4\mu ^{3}h^{3}+\left(
J_{1}^{2}+J_{2}^{2}+J_{3}^{2}\right) \mu h}{J_{1}J_{2}J_{3}},
\label{lambda3}
\end{equation}%
with $h_{c1}$, $h_{c2}$ and $h_{c3}$ given in the eq.(\ref{criticalfields3}%
), and finally for $n=4$ by.%
\begin{equation}
M^{z}=\left\{
\begin{array}{lll}
\frac{\mu }{2} &  & h\geq h_{c1}\equiv h_{s}, \\
\frac{\mu }{2}-\frac{\mu }{4\pi }\arccos \Lambda _{4} &  &
h_{c2}\leq h\leq
h_{c1}, \\
\frac{\mu }{4}, &  & h_{c3}\leq h\leq h_{c2}, \\
\frac{\mu }{4}-\frac{\mu }{4\pi }\arccos (-\Lambda _{4}) &  &
h_{c4}\leq
h\leq h_{c3}, \\
0 &  & h\leq h_{c4},%
\end{array}%
\right. \text{ }  \label{tauz4}
\end{equation}%
where $h_{c1}$, $h_{c2}$, $h_{c3}$ and $h_{c4\text{ }}$ are given in eq.(\ref%
{criticalfields4}), and
\begin{equation}
\Lambda _{4}=\frac{8\mu ^{4}h^{4}-2\left(
J_{3}^{2}+J_{2}^{2}+J_{4}^{2}+J_{1}^{2}\right) \mu ^{2}h^{2}}{%
J_{1}J_{2}J_{3}J_{4}}+\frac{\left(
J_{1}^{2}J_{3}^{2}+J_{2}^{2}J_{4}^{2}\right)
}{2J_{1}J_{2}J_{3}J_{4}}.
\end{equation}

For arbitrary $n,$ between two critical fields,\ namely, $h_{cj}$ and $%
h_{cj+1},$ with $h_{c1}\equiv h_{s}$, a general expression can be
obtained
for $M_{j}^{z}$ as a function of $h$ and is given by%
\begin{equation}
M_{j}^{z}=M_{pj}^{z}-\frac{\mu }{n\pi }\arccos \left[ \left( -1\right) ^{n+%
\frac{j+3}{2}}\Lambda _{n}\right] ,  \label{maggeral}
\end{equation}%
where $M_{pj}^{z}=\mu \left[ n-2(j-1)\right] /2n$ is the
magnetization in
the upper plateau and $\Lambda _{n}$ is given by%
\begin{equation}
\Lambda _{n}=\frac{\Im (0,h)}{2J_{1}J_{2}...J_{n}}.  \label{lambdan}
\end{equation}

Since the model has azimuthal symmetry, for homogeneous magnetic
moments, the values of the induced magnetization per site at the
plateaus, $M_{p},$ satisfy the quantization condition
\cite{oshikawa:1997}
\begin{equation}
n(\mu /2-M_{p}^{z})=\mu \times integer, \label{szquantization}
\end{equation}%
which is verified by the results shown above in eqs.(\ref{tauz2}-\ref{tauz4}%
). For different magnetic moments,\ although the azimuthal symmetry
is preserved, this condition is no more satisfied since the total
magnetization operator does not commute with the Hamiltonian.
However, it must be noted that eq.(\ref{szquantization}) is always
satisfied provided we replace the induced magnetization per site by
the average spin component in the field per site and assume $\mu
=1.$

For different $\mu ^{\prime }s,$ when there are no magnetization
plateaus, we can define an order parameter associated with each
second order quantum transition given similarly to the case where we
have identical $\mu ^{\prime }s$ \cite{delima:2002} as
\begin{equation}
\widetilde{M}^{z}=\left\vert M^{z}-M_{c}^{z}\right\vert ,
\end{equation}%
where $M_{c}^{z}$ is the magnetization at the critical point, and
from this we can show that the critical exponent $\beta ,$ obtained
from the scaling relation
\begin{equation}
\widetilde{M}^{z}\sim \left\vert h-h_{c}\right\vert ^{\beta },
\label{beta}
\end{equation}%
is equal to $1/2,$ and the exponent $\gamma $ associated with the
isothermal susceptibility$,$ obtained from
\begin{equation}
\chi _{T}^{zz}\sim \left\vert h-h_{c}\right\vert ^{-\gamma },
\label{quizscaling}
\end{equation}%
is also equals to $1/2$.

For the special case when the $J\prime s$ satisfy the special condition, eq( %
\ref{specialcondition}), the critical exponents $\beta $ and $\gamma
$ associated with the transition that occurs at $h=0$ are given by 1
and 0 respectively. Therefore, the quantum transitions of the
inhomogeneous model, apart from this special point, belong to the
same universality class as the ones in the homogeneous model
\cite{niemeijer:1967}.

\section{Static and dynamic correlations $\left\langle \protect\tau _{l}^{z}%
\protect\tau _{l+r}^{z}\right\rangle $}

As it has been shown
\cite{siskens:1974,capel:1977,goncalvestese:1977}, in the
thermodynamic limit, the correlation function $\left\langle
S_{l,m}^{z}(t)S_{l+r,m^{\prime }}^{z}(0)\right\rangle $ can be
obtained from
the expression%
\begin{equation}
\left\langle S_{l,m}^{z}(t)S_{l+r,m^{\prime }}^{z}(0)\right\rangle =\frac{Tr%
\left[ \exp (-\beta H^{-})\exp (iH^{-}t)S_{l,m}^{z}\exp
(-iH^{-}t)S_{l+r,m^{\prime }}^{z}\right] }{Tr\left[ \exp (-\beta H^{-})%
\right] },
\end{equation}%
and consequently we can express this correlation in terms of fermion
operators. Therefore, the dynamic correlation between the effective
spins in the field direction
\begin{equation}
\left\langle \tau _{l}^{z}(t)\tau _{l+r}^{z}(0)\right\rangle =\frac{1}{n^{2}}%
\sum_{m,m^{\prime }=1}^{n}\mu _{m}\mu _{m^{\prime }}\left\langle
S_{l,m}^{z}(t)S_{l+r,m^{\prime }}^{z}(0)\right\rangle ,
\end{equation}%
can be written as
\begin{eqnarray}
\left\langle \tau _{l}^{z}(t)\tau _{l+r}^{z}(0)\right\rangle &=&\frac{1}{%
n^{2}}\left\{ \sum_{m,m^{\prime }=1}^{n}\mu _{m}\mu _{m^{\prime
}}\left\langle c_{l,m}^{\dagger }(t)c_{l,m}(t)c_{l+r,m^{\prime
}}^{\dagger
}(0)c_{l+r,m^{\prime }}(0)\right\rangle \right. -  \notag \\
&-&\frac{1}{2}\mu _{m}\mu _{m^{\prime }}\left( \left\langle
c_{l,m}^{\dagger }(t)c_{l,m}(t)\right\rangle +\left\langle
c_{l+r,m^{\prime }}^{\dagger
}(0)c_{l+r,m^{\prime }}(0)\right\rangle \right) +  \notag \\
&&+\left. \frac{\mu _{m}\mu _{m^{\prime }}}{4}\right\} ,
\end{eqnarray}%
and by using Wick's theorem \cite{parry:1973}, we obtain
\begin{equation}
\left\langle \tau _{l}^{z}(t)\tau _{l+r}^{z}(0)\right\rangle =\frac{1}{n^{2}}%
\sum_{m,m^{\prime }=1}^{n}\mu _{m}\mu _{m^{\prime }}\left\langle
c_{l,m}^{\dagger }(t)c_{l+r,m^{\prime }}(0)\right\rangle
\left\langle c_{l,m}(t)c_{l+r,m^{\prime }}^{\dagger
}(0)\right\rangle +\left\langle \tau _{l}^{z}\right\rangle ^{2}.
\end{equation}%
From eqs.(\ref{inversetransformation}), we can write the
contractions
\begin{eqnarray}
\left\langle c_{l,m}^{\dagger }(t)c_{l^{\prime },m^{\prime
}}(0)\right\rangle &=&\frac{1}{N}\sum\limits_{q,k}e^{iqd(l-l^{\prime
})}u_{q,km}^{\ast }u_{q,km^{\prime }}n_{q,k}\exp (i\varepsilon
_{q,k}t),
\notag \\
\left\langle c_{l,m}(t)c_{l^{\prime },m^{\prime }}^{\dagger
}(0)\right\rangle &=&\frac{1}{N}\sum\limits_{q,k}e^{-iqd(l-l^{\prime
})}u_{q,km}u_{q,km^{\prime }}^{\ast }\left( 1-n_{q,k}\right) \exp
(-i\varepsilon _{q,k}t),  \label{contractions}
\end{eqnarray}%
and from this equation, we obtain%
\begin{eqnarray}
\left\langle \tau _{l}^{z}(t)\tau _{l+r}^{z}(0)\right\rangle &=&\frac{1}{%
n^{2}N^{2}}\sum_{q,k,m}\sum_{q^{\prime },k^{\prime },m^{\prime }}\mu
_{m}\mu _{m^{\prime }}u_{q,km}^{\ast }u_{q,km^{\prime }}u_{q^{\prime
},k^{\prime
}m}u_{q^{\prime },k^{\prime }m^{\prime }}^{\ast }\times  \notag \\
&\times &n_{q,k}\left( 1-n_{q^{\prime },k^{\prime }}\right)
e^{-idr(q-q^{\prime })}e^{i\left( \varepsilon _{q,k}-\varepsilon
_{q^{\prime },k^{\prime }}\right) t}+\left\langle \tau
_{l}^{z}\right\rangle ^{2}, \label{dynamiccorrelationT>0}
\end{eqnarray}%
which in the limit $T\rightarrow 0$ can be written as
\begin{eqnarray}
\left\langle \tau _{l}^{z}(t)\tau _{l+r}^{z}(0)\right\rangle &=&\frac{1}{%
n^{2}N^{2}}\sum_{\substack{ q,k  \\ (for\text{ }\varepsilon _{q,k}\leq 0)%
\text{ }}}\sum_{\substack{ q^{\prime },k^{\prime }  \\ (for\text{ }%
\varepsilon _{q^{\prime },k^{\prime }}\geq 0)}}\sum_{m,m^{\prime
}}\mu
_{m}\mu _{m^{\prime }}u_{q,km}^{\ast }u_{q,km^{\prime }}\times  \notag \\
&&\times u_{q^{\prime },k^{\prime }m}u_{q^{\prime },k^{\prime
}m^{\prime }}^{\ast }e^{-idr(q-q^{\prime })}e^{i\left( \varepsilon
_{q,k}-\varepsilon _{q^{\prime },k^{\prime }}\right) t}+\left\langle
\tau _{l}^{z}\right\rangle ^{2}.  \label{dynamiccorrelationT=0}
\end{eqnarray}%
The static correlation $\left\langle \tau _{l}^{z}\tau
_{l+r}^{z}\right\rangle $ is immediately obtained from the dynamic
one by making $t=0$ and we obtain, for finite $T,$
\begin{eqnarray}
\left\langle \tau _{l}^{z}\tau _{l+r}^{z}\right\rangle &=&\frac{1}{n^{2}N^{2}%
}\sum_{q,m,k}\sum_{q^{\prime },m^{\prime },k^{\prime }}\mu _{m}\mu
_{m^{\prime }}u_{q,km}^{\ast }u_{q,km^{\prime }}u_{q^{\prime
},k^{\prime
}m}u_{q^{\prime },k^{\prime }m^{\prime }}^{\ast }\times  \notag \\
&\times &n_{q,k}\left( 1-n_{q^{\prime },k^{\prime }}\right)
e^{-idr(q-q^{\prime })}+\left\langle \tau _{l}^{z}\right\rangle
^{2},
\end{eqnarray}%
and
\begin{gather}
\left\langle \tau _{l}^{z}\tau _{l+r}^{z}\right\rangle =\frac{1}{n^{2}N^{2}}%
\sum_{\substack{ q,k  \\ (for\text{ }\varepsilon _{q,k}\leq 0)\text{
}}}\sum _{\substack{ q^{\prime },k^{\prime }  \\ (for\text{
}\varepsilon _{q^{\prime },k^{\prime }}\geq 0)}}\sum_{m,m^{\prime
}}\mu _{m}\mu _{m^{\prime
}}u_{q,km}^{\ast }u_{q,km^{\prime }}\times  \notag \\
\times u_{q^{\prime },k^{\prime }m}u_{q^{\prime },k^{\prime
}m^{\prime }}^{\ast }e^{-idr(q-q^{\prime })}+\left\langle \tau
_{l}^{z}\right\rangle ^{2},
\end{gather}%
in the limit $T\rightarrow 0.$

For different $\mu ^{\prime }s,$ we present in Fig. 5, for a chain with $%
n=8, $ at $T=0$, the static correlation as a function of the
distance between cells for field values above, below and at a
critical point. As expected, for fields below the critical one,
which in this case corresponds to the region which contains the
divergence of the isothermal susceptibility, the correlation
presents an oscillatory behaviour. As in the alternating
superlattice \cite{delima:2002}, the period of the oscillation
corresponds to the correlation length and goes to infinity as we
approach the critical field, and there is no oscillation in the
correlation for fields in the disordered region.

The real and imaginary parts of the dynamic correlation as function
of the time, for a chain with the same set of parameters of Fig. 5,
are presented in Fig. 6. Differently from the case where we have
identical $\mu ^{\prime }s,$ in the region between two critical
fields, where the isothermal susceptibility is finite, the dynamic
correlation varies with the field, and for $h\geqslant h_{s}$ is
time-independent and equal to $\left\langle \tau
_{l}^{z}\right\rangle ^{2}$.

\section{Dynamic susceptibility $\protect\chi _{q}^{zz}(\protect\omega )$}

Introducing the time Fourier transform in the dynamic correlation $%
\left\langle \tau _{l}^{z}(t)\tau _{l+r}^{z}(0)\right\rangle $,%
\begin{equation}
\left\langle \tau _{l}^{z}\tau _{l+r}^{z}\right\rangle _{\omega
}\equiv \frac{1}{2\pi }\int_{-\infty }^{\infty }\left\langle \tau
_{l}^{z}(t)\tau _{l+r}^{z}(0)\right\rangle e^{i\omega t}dt,
\label{transf-fourier-temp}
\end{equation}%
we obtain from eq.(\ref{dynamiccorrelationT>0}) the result
\begin{eqnarray}
\left\langle \tau _{l}^{z}\tau _{l+r}^{z}\right\rangle _{\omega } &=&\frac{1%
}{n^{2}N^{2}}\sum_{q,k,m}\sum_{q^{\prime },k^{\prime },m^{\prime
}}\mu _{m}\mu _{m^{\prime }}u_{q,km}^{\ast }u_{q,km^{\prime
}}u_{q^{\prime },k^{\prime }m}u_{q^{\prime },k^{\prime }m^{\prime
}}^{\ast }n_{q,k}\times
\notag \\
&&\times \left( 1-n_{q^{\prime },k^{\prime }}\right)
e^{-idr(q-q^{\prime })}\delta \left( \omega -\varepsilon
_{q,k}-\varepsilon _{q^{\prime },k^{\prime }}\right) +\left\langle
\tau _{l}^{z}\right\rangle ^{2}\delta (\omega ).
\label{corr-fourier-temp}
\end{eqnarray}%
By considering the spacial Fourier transform in the field direction,
\begin{equation}
\left\langle \tau _{q}^{z}\tau _{-q}^{z}\right\rangle _{\omega
}\equiv \sum_{r}\left\langle \tau _{l}^{z}\tau
_{l+r}^{z}\right\rangle _{\omega }e^{idrq},
\label{transf-fourier-esp}
\end{equation}%
we obtain the correlation as function of $\omega $ and $q$, in the
form
\begin{eqnarray}
\left\langle \tau _{q}^{z}\tau _{-q}^{z}\right\rangle _{\omega } &=&\frac{1}{%
n^{2}N}\sum_{q^{\prime }}\sum_{k,m}\sum_{k^{\prime },m^{\prime }}\mu
_{m}\mu _{m^{\prime }}u_{q^{\prime },km}^{\ast }u_{q^{\prime
},km^{\prime }}u_{q^{\prime }-q,k^{\prime }m}u_{q^{\prime
}-q,k^{\prime }m^{\prime
}}^{\ast }\times  \notag \\
&\times &n_{q^{\prime },k}\left( 1-n_{q^{\prime }-q,k^{\prime
}}\right) \delta \left( \omega +\varepsilon _{q^{\prime
},k}-\varepsilon _{q^{\prime
}-q,k^{\prime }}\right) +  \notag \\
&+&\left\langle \tau _{l}^{z}\right\rangle ^{2}\delta (\omega
)\delta _{q,0}.
\end{eqnarray}%
From this expression we can determine the dynamic susceptibility
$\chi _{q}^{zz}(\omega ),$ which is given by\cite{zubarev:1960}
\begin{equation}
\chi _{q}^{zz}(\omega )\equiv -\ 2\pi \left\langle \left\langle \tau
_{q}^{z};\tau _{-q}^{z}\right\rangle \right\rangle ,
\label{susdin1}
\end{equation}%
where the Green function $\left\langle \left\langle \tau
_{q}^{z};\tau _{-q}^{z}\right\rangle \right\rangle $ is obtained
from
\begin{equation}
\left\langle \left\langle \tau _{q}^{z};\tau _{-q}^{z}\right\rangle
\right\rangle =\frac{1}{2\pi }\int_{-\infty }^{\infty }\frac{\left(
1-e^{-\beta \omega }\right) \left\langle \tau _{q}^{z}\tau
_{-q}^{z}\right\rangle }{\omega -\omega ^{\prime }}d\omega ^{\prime
}, \label{susdin2}
\end{equation}%
and we obtain
\begin{eqnarray}
\chi _{q}^{zz}(\omega ) &=&-\frac{1}{n^{2}N}\sum_{q^{\prime
}}\sum_{k,m}\sum_{k^{\prime },m^{\prime }}\mu _{m}\mu _{m^{\prime
}}u_{q^{\prime },km}^{\ast }u_{q^{\prime },km^{\prime }}u_{q^{\prime
}-q,k^{\prime }m}u_{q^{\prime }-q,k^{\prime }m^{\prime }}^{\ast
}\times
\notag \\
&\times &\frac{n_{q^{\prime },k}-n_{q^{\prime }-q,k^{\prime
}}}{\omega +\varepsilon _{q^{\prime },k}-\varepsilon _{q^{\prime
}-q,k^{\prime }}}, \label{dynamicsusceptibility}
\end{eqnarray}%
where we have used the identity
\begin{equation}
1-n_{q,k}=e^{\beta \varepsilon _{q,k}}n_{q,k}.
\label{ident-num-fermi}
\end{equation}

We can also write $\chi _{q}^{zz}(\omega )$ in the form%
\begin{eqnarray}
\chi _{q}^{zz}(\omega ) &=&\frac{1}{n^{2}2N}\sum_{q^{\prime
}}\sum_{k,m}\sum_{k^{\prime },m^{\prime }}\mu _{m}\mu _{m^{\prime
}}u_{q^{\prime },km}^{\ast }u_{q^{\prime },km^{\prime }}u_{q^{\prime
}-q,k^{\prime }m}u_{q^{\prime }-q,k^{\prime }m^{\prime }}^{\ast
}\times
\notag \\
&\times &\frac{\tanh (\beta \varepsilon _{q^{\prime },k}/2)-\tanh
(\beta \varepsilon _{q^{\prime }-q,k^{\prime }}/2)}{\omega
+\varepsilon _{q^{\prime },k}-\varepsilon _{q^{\prime }-q,k^{\prime
}}},  \label{susc-omega-q}
\end{eqnarray}%
and from this obtain the static susceptibility $\chi _{0}^{zz}(0)$
which is
explicitly given by%
\begin{gather}
\chi _{0}^{zz}(0)=\frac{1}{n^{2}2N}\sum_{q^{\prime }}\sum_{\substack{ %
k,k^{\prime }  \\ k\neq k^{\prime }}}\sum_{m,m^{\prime }}\mu _{m}\mu
_{m^{\prime }}u_{q^{\prime },km}^{\ast }u_{q^{\prime },km^{\prime
}}u_{q^{\prime },k^{\prime }m}u_{q^{\prime },k^{\prime }m^{\prime
}}^{\ast
}\times  \notag \\
\times \frac{\tanh (\beta \varepsilon _{q^{\prime },k}/2)-\tanh
(\beta \varepsilon _{q^{\prime },k^{\prime }}/2)}{\varepsilon
_{q^{\prime
},k}-\varepsilon _{q^{\prime },k^{\prime }}}+  \notag \\
+\frac{\beta }{n^{2}4N}\sum_{q^{\prime }k}\sum_{m,m^{\prime }}\mu
_{m}\mu _{m^{\prime }}u_{q^{\prime },km}^{\ast }u_{q^{\prime
},km^{\prime
}}u_{q^{\prime },km}u_{q^{\prime },km^{\prime }}^{\ast }\text{sech}%
^{2}(\beta \varepsilon _{q^{\prime },k}/2).
\end{gather}%
As in the uniform model \cite{katsura:1970}, it can be shown from
the previous expression that $\chi _{0}^{zz}(0)$ is equal to the
isothermal
susceptibility $\chi _{T}^{zz}$ given in eq.(\ref{susceptibilityisothermal})$%
.$

The real and imaginary parts of $\chi _{q}^{zz}(\omega ),$ as usual,
are
obtained by considering $\chi _{q}^{zz}(\omega -i\epsilon )$ in the limit $%
\epsilon \rightarrow 0$ in eq.(\ref{susc-omega-q}) and are given by%
\begin{eqnarray}
Re\chi _{q}^{zz}(\omega ) &=&\frac{1}{n^{2}2N}P\sum_{q^{\prime
}}\sum_{k,m}\sum_{k^{\prime },m^{\prime }}\mu _{m}\mu _{m^{\prime
}}u_{q^{\prime },km}^{\ast }u_{q^{\prime },km^{\prime }}u_{q^{\prime
}-q,k^{\prime }m}u_{q^{\prime }-q,k^{\prime }m^{\prime }}^{\ast
}\times
\notag \\
&&\times \frac{\tanh (\beta \varepsilon _{q^{\prime },k}/2)-\tanh
(\beta \varepsilon _{q^{\prime }-q,k^{\prime }}/2)}{\omega
+\varepsilon _{q^{\prime },k}-\varepsilon _{q^{\prime }-q,k^{\prime
}}},
\end{eqnarray}%
where $P$ denotes Cauchy principal value, and%
\begin{eqnarray}
Im\chi _{q}^{zz}(\omega ) &=&\frac{\pi }{n^{2}2N}\sum_{q^{\prime
}}\sum_{k,m}\sum_{k^{\prime },m^{\prime }}\mu _{m}\mu _{m^{\prime
}}u_{q^{\prime },km}^{\ast }u_{q^{\prime },km^{\prime }}u_{q^{\prime
}-q,k^{\prime }m}u_{q^{\prime }-q,k^{\prime }m^{\prime }}^{\ast
}\times
\notag \\
&&\times (\tanh (\beta \varepsilon _{q^{\prime },k}/2)-\tanh (\beta
\varepsilon _{q^{\prime }-q,k^{\prime }}/2))\times  \notag \\
&&\times \delta (\omega +\varepsilon _{q^{\prime },k}-\varepsilon
_{q^{\prime }-q,k^{\prime }}).
\end{eqnarray}%
At $T=0,$ these results, as function of $\omega ,$ are shown in Fig. 7, for $%
n=4$ and different wave-vectors for a field in the disordered
region. For any wave-vector there are the same number of bands in
the imaginary part of the susceptibility, and this number depends on
the size of the unit cell and on the values of the parameters.

In Fig. 8 it is shown the static susceptibility, $\chi _{q}^{zz}(0),$ at $%
T=0,$\textbf{\ }for\ the same lattice parameters of Fig. 7, as a
function of the field. For $q=0$, since it is identical to the
isothermal susceptibility, it diverges at the critical fields. The
singularities at non-zero and at the Brillouin zone boundary
wave-vectors can be associated with critical points, whereas the
ones for different wave-vectors are related to oscillations of the
spin correlations and can also be associated with the unstable
critical points present in the study of the system within the real
space renormalization group \cite{jullien:1979}.

\section{ T=0 isothermal and dynamic susceptibilities $\protect\chi %
_{T}^{xx} $ and $\protect\chi _{q}^{xx}(\protect\omega )$}

At $T=0$, and for h$\geq h_{s},$ we can obtain from the eq.(\ref%
{correlationxx}) the time Fourier transform of dynamic correlation $%
\left\langle S_{1,m}^{x}(t)S_{1+r,m^{\prime }}^{x}(0)\right\rangle
,$ which
is given by%
\begin{equation}
\left\langle S_{1,m}^{x}S_{1+r,m^{\prime }}^{x}\right\rangle _{\omega }=%
\frac{1}{2\pi }\int_{-\infty }^{\infty }\exp (i\omega t)\left\langle
S_{1,m}^{x}(t)S_{1+r,m^{\prime }}^{x}(0)\right\rangle dt,
\end{equation}%
and can be written as
\begin{equation}
\left\langle S_{1,m}^{x}S_{1+r,m^{\prime }}^{x}\right\rangle _{\omega }=%
\frac{(-1)^{nr+m^{\prime }-m}}{4N}\sum_{q,k}\exp
(-iqdr)u_{q,km}^{\ast }u_{qkm^{\prime }}\delta (\omega +\varepsilon
_{q,k}).  \label{corrxxomega}
\end{equation}%
Following Gon\c{c}alves \cite{goncalves:1986}, we can write the
local
dynamic susceptibility $\chi ^{xx}(\omega ,r,m,m^{\prime })$ as%
\begin{equation}
\chi ^{xx}(\omega ,r,m,m^{\prime })=-\lim_{\beta \rightarrow \infty
}\int_{-\infty }^{\infty }\frac{\left[ 1-\exp (-\beta \omega
)\right]
\left\langle S_{1,m}^{x}S_{1+r,m^{\prime }}^{x}\right\rangle _{\omega }}{%
\omega -\omega ^{\prime }}d\omega ^{\prime },
\end{equation}%
since $\varepsilon _{q,k}\leq 0,$ and by using
eq.(\ref{corrxxomega}) we
obtain%
\begin{equation}
\chi ^{xx}(\omega ,r,m,m^{\prime })=-\frac{\left( -1\right)
^{nr+m^{\prime
}-m}}{4N}\sum_{q,k}\frac{\exp (-iqdr)u_{q,km}^{\ast }u_{qkm^{\prime }}}{%
\omega +\varepsilon _{q,k}}.  \label{localsusceptibility}
\end{equation}%
The susceptibility associated with the cell effective spin in the
$x$
direction is given by%
\begin{equation}
\chi ^{xx}(\omega ,r)=\sum_{m,m^{\prime }}\chi ^{xx}(\omega
,r,m,m^{\prime }),
\end{equation}%
and by using eq.(\ref{localsusceptibility}) we can write
\begin{equation}
\chi ^{xx}(\omega ,r)=-\frac{\left( -1\right) ^{nr}}{4N}\sum_{q,k,m,m^{%
\prime }}\left( -1\right) ^{m^{\prime }-m}\frac{\exp
(-iqdr)u_{q,km}^{\ast }u_{q,km^{\prime }}}{\omega +\varepsilon
_{q,k}}
\end{equation}%
and from this the final result%
\begin{equation}
\chi _{q}^{xx}(\omega )=-\sum_{k,m,m^{\prime }}\frac{\left(
-1\right)
^{m-m^{\prime }}u_{q^{\prime },km}^{\ast }u_{q^{\prime },km^{\prime }}}{%
4\left( \omega +\varepsilon _{q^{\prime },k}\right) },
\label{suscepdinamicaxx}
\end{equation}%
where%
\begin{equation}
q^{\prime }=\left\{
\begin{array}{c}
q\text{ \ \ \ \ \ \ \ \ \ for }n\text{ even, } \\
q-\frac{\pi }{d}\text{ \ \ \ for }n\text{ odd.}%
\end{array}%
\right.
\end{equation}%
From eq.(\ref{suscepdinamicaxx}) by considering the $\chi
_{q}^{xx}(\omega -i\epsilon )$ in the limit $\epsilon \rightarrow 0$
we can obtain
immediately the real and imaginary parts which are given by%
\begin{equation}
Re\chi _{q}^{xx}(\omega )=-P\sum_{k,m,m^{\prime }}\frac{\left(
-1\right) ^{m-m^{\prime }}u_{q^{\prime },km}^{\ast }u_{q^{\prime
},km^{\prime }}}{4\left( \omega +\varepsilon _{q^{\prime },k}\right)
},
\end{equation}%
where $P$ denotes Cauchy principal value, and%
\begin{equation}
Im\chi _{q}^{xx}(\omega )=-\frac{\pi }{4}\sum_{k,m,m^{\prime
}}\left( -1\right) ^{m-m^{\prime }}u_{q^{\prime },km}^{\ast
}u_{q^{\prime },km^{\prime }}\delta \left( \omega +\varepsilon
_{q^{\prime },k}\right) .
\end{equation}%
We can also obtain, in the limit considered, the local isothermal
susceptibility from the expression \cite{horiguchi:1975}%
\begin{equation}
\chi _{T}^{xx}(r,m,m^{\prime })=\lim_{\beta \rightarrow \infty
}\int_{0}^{\beta }\left\langle S_{1,m}^{x}(-i\lambda
)S_{1+r,m^{\prime }}^{x}(0)\right\rangle d\lambda ,
\end{equation}%
which, from eq.(\ref{correlationxx}), can be written as%
\begin{equation}
\chi _{T}^{xx}(r,m,m^{\prime })=-\frac{\left( -1\right) ^{nr+m^{\prime }-m}}{%
4N}\sum_{q,k}\frac{\exp (-iqdr)u_{q,km}^{\ast }u_{q,km^{\prime }}}{%
\varepsilon _{q,k}}.
\end{equation}%
From this result we can immediately conclude that the isothermal
susceptibility, $\chi _{T}^{xx}=\sum_{r,m,m^{\prime }}\chi
_{T}^{xx}(r,m,m^{\prime }),$ is equal to the static one, $\chi
_{q}^{xx}(\omega ),$ in the limit $q\rightarrow 0.$

For equal magnetic moments, n=8 and the exchange parameters of Fig.
3(a), the real and imaginary parts of the dynamic correlations
$\left\langle S_{1,m}^{x}(t)S_{1+r,m^{\prime }}^{x}(0)\right\rangle
$ and $\left\langle S_{1,m}^{x}(t)S_{1+r,m^{\prime
}}^{x}(0)\right\rangle $ are shown in Figs. 9 and 10 respectively,
for various magnetic fields. For large $t,$ these correlations
factorize and tend to zero, since the average $\left\langle
S_{l,m}^{x}\right\rangle \rightarrow 0.$

Due to the simple form of the dynamic susceptibility, eq.(\ref%
{suscepdinamicaxx} ), it presents poles of order one at the
excitation energies. This means that the imaginary part corresponds
to a set of delta functions for a given wave-vector.

In Fig. 11 we present the static susceptibility $\chi
_{q}^{xx}(\omega ),$ at $T=0,$ as a function of the field for a
chain with $n=8$. For the ferromagnetic case shown in Fig. 11(a) it
diverges at the critical field for $q=0$ only. On the other hand, if
we change the sign of a single exchange interaction, it diverges for
$q$ at the zone boundary, namely, $q=\pi /8.$ This is shown in Fig.
11(b), and means that the system in terms of cells has an
antiferromagnetic behaviour. Although it is not presented, it can be
shown that the order ferro (antiferro) is associated to the sign +
(-) of the product of the exchange interactions within the cells.

\section{Conclusions}

We have considered in this work the isotropic XY model on the
inhomogeneous periodic chain with $N$ cells, $n$ sites per cell. The
exact solution has been obtained in the general case, where we have
$n$ different exchange constants and magnetic moments, and for
arbitrary temperatures.

At $T=0$, for equal magnetic moments, we have shown that the induced
magnetization, as a function of the field, presents plateaus which
satisfy the quantization condition shown in eq.
(\ref{szquantization}), and that the isothermal susceptibility $\chi
_{T}^{zz}$ is zero within the plateaus and diverges at the limiting
fields which characterize the multiple second order quantum
transitions. On the other hand, f or different magnetic moments,
although the system presents multiple quantum transitions, the
induced magnetization plateaus are suppressed and the isothermal
susceptibility, which also diverges at the critical points, is
always different from zero. In both cases, the disordered phases
correspond to the gaps in the excitation spectrum and to the regions
where $\chi _{T}^{zz}$ is finite, which are identical to the
plateaus for the case of identical magnetic moments. For $n=1,2,3$,
explicit analytical expressions are presented for the critical
fields, in the general case, and for the induced magnetization when
we have identical magnetic moments.

The number of transitions, for different $\mu ^{\prime }s,$ is
always equal to $n,$ even for the special case where the exchange
constants, for $n$ even, satisfy the condition presented in
eq.(\ref{specialcondition}), where at zero field there is no gap
between the ground state and the first excited state. It should be
noted that in this case, $\chi _{T}^{zz}$ diverges at both sides of
the zero field transition\ and that, for identical $\mu ^{\prime
}s,$ this transition is suppressed.

The critical exponents are identical to those of the uniform model,
$\alpha
=1/2,$ $\beta =1/2,$ $\gamma =1/2,$ and for the special point at zero field $%
\alpha =0,$ $\beta =1,$ $\gamma =0,$ and naturally they satisfy the
Rushbrook relation $\alpha +$ 2$\beta +$ $\gamma =2$ (see e.g. ref.\cite%
{stanley:1971}). It should be noted that in these transitions
$\alpha =\gamma $ \cite{delima:1997}.

The correlation length, which corresponds to the period of the
oscillation of the static correlation, $\left\langle \tau
_{l}^{z}\tau _{l+r}^{z}\right\rangle ,$ as in the homogeneous chain
and alternating superlattice \cite{delima:2002}$,$ diverges at the
critical points irrespective of the values of the magnetic moments.
On the other hand, the dynamic correlation $\left\langle \tau
_{l}^{z}(t)\tau _{l+r}^{z}(0)\right\rangle ,$ which is independent
of the field in the plateaus regions for equal magnetic moments, is
field dependent when we have different magnetic moments.

The isothermal susceptibility $\chi _{T}^{zz}$ is equal to the the
static one, $\chi _{0}^{zz}(0),$ at any temperature. Independently
of the value of the field and of the magnetic moments, the imaginary
part of the dynamic susceptibility $\chi _{q}^{zz}(\omega )$
presents several bands whose number does not depend on $q$. As in
the alternating superlattice, the discontinuities of the imaginary
part at the band edges correspond to singularities in the real part.
Associated with the critical behaviour are the descontinuities which
occur at zero or at the Brillouin zone boundary wave-vectors,
whereas those that occur for other values of the wave-vectors are
related to unstable critical points \cite{jullien:1979}.

At $T=0$ and for fields greater than the saturation field, we
obtained the dynamic correlation $\left\langle
S_{1,m}^{x}(t)S_{1+r,m^{\prime }}^{x}(0)\right\rangle $ and the
isothermal and wave-vector dynamic susceptibilities in the $x$
direction. We have shown that the isothermal susceptibility \ $\chi
_{T}^{xx}$ is identical to the static one $\chi _{0}^{zz}(0),$ and
that in terms of the magnetization of the cells the system presents
a ferromagnetic or antiferromagnetic order depending on the sign of
the product of the exchange interactions within the cells.

\textbf{Acknowledgements}

The authors would like to thank the Brazilian agencies CNPq, Capes
and Finep for partial financial support. They would also like to
thank Prof. S. R. A.
Salinas for his hospitality during their stay at the Departamento de F\'{\i}%
sica Geral, Instituto de F\'{\i}sica, Universidade de S\~{a}o Paulo,
and Dr. A. P. Vieira for valuable discussions.

\bigskip

\bigskip

\bigskip

\bigskip

\bigskip

\bigskip

\bigskip

\bigskip \bigskip

\pagebreak
\begin{figure}
\begin{center}
\includegraphics[width=14cm]{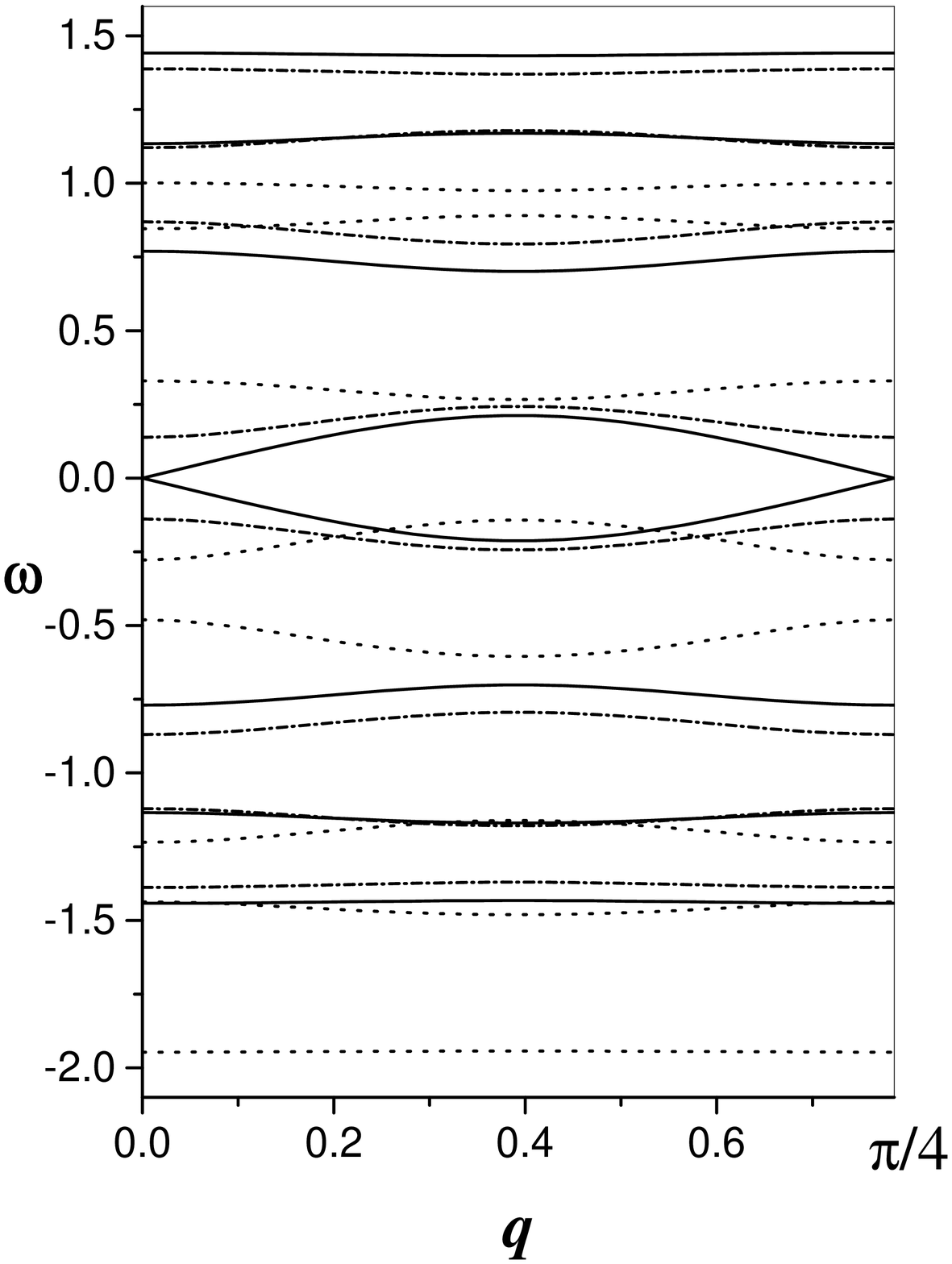}
\end{center}
\caption{Excitation spectrum for $n=8$%
, $J_{1}=1,$ $J_{2}=2,$ $J_{3}=1.5,$ $J_{4}=5/3,$ $J_{5}=2/3,$ $J_{6}=3/5,$ $%
J_{7}=2,$ $J_{8}=1$ for $\protect\mu _{1}=....=\protect\mu _{8}=1$
and $h=0$
(continuous line) and for $\protect\mu _{1}=3,$ $\protect\mu _{2}=4,$ $%
\protect\mu _{3}=\protect\mu _{5}=\protect\mu _{7}=1.5,$ $\protect\mu %
_{4}=2.5,$ $\protect\mu _{6}=\protect\mu _{8}=1$ and $h=0.2$ (dotted
line),
and for $J_{1}=2,$ $J_{2}=1,$ $J_{3}=1.5,$ $J_{4}=5/3,$ $J_{5}=2/3,$ $%
J_{6}=3/5,$ $J_{7}=2,$ $J_{8}=1,$ $\protect\mu _{1}=....=\protect\mu
_{8}=1$ and $h=0$ (dot-dashed line).} \label{fig2}
\end{figure}

\pagebreak
\begin{figure}
\begin{center}
\includegraphics[width=14cm]{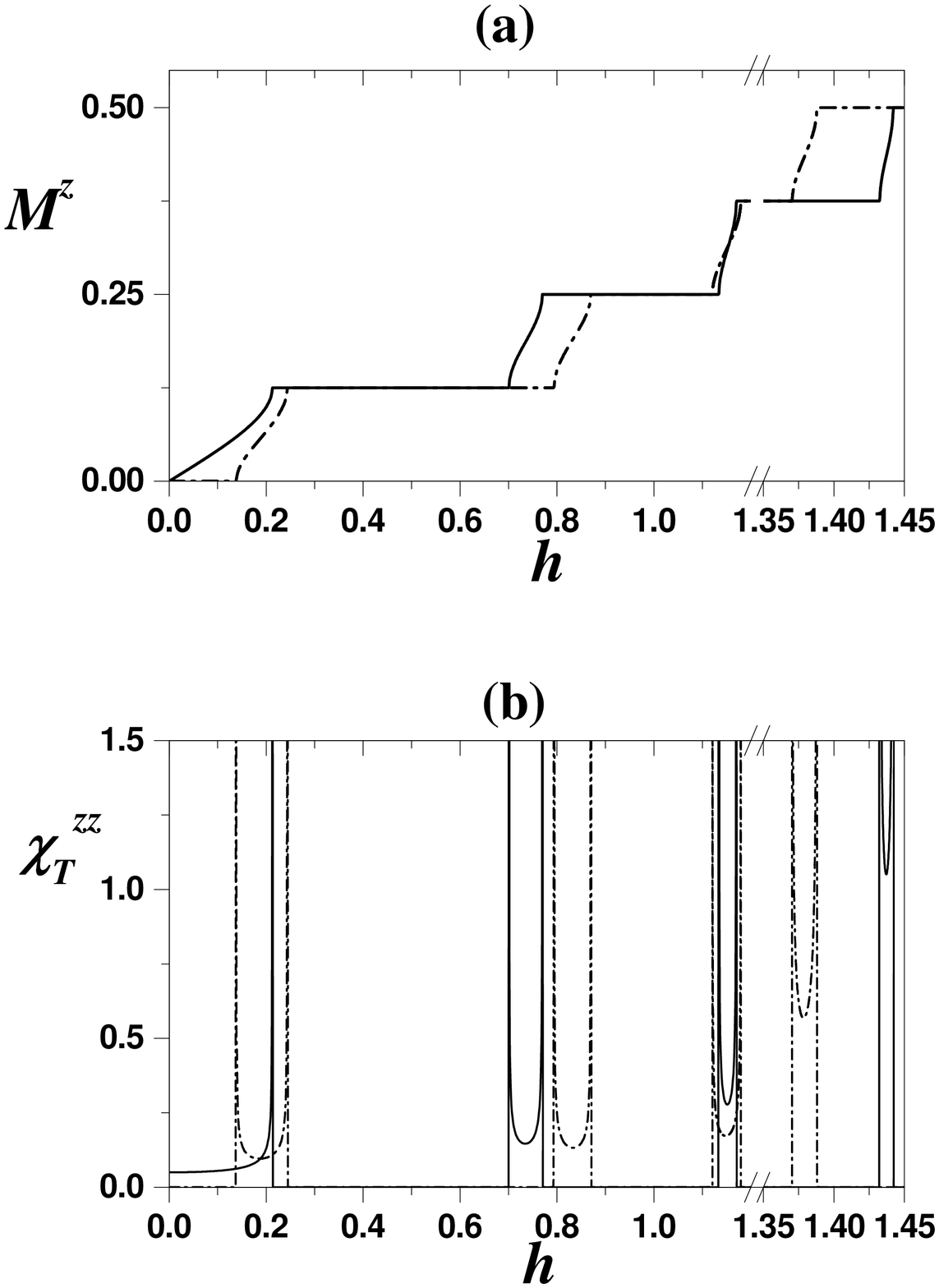}
\end{center}
\caption{(a) Magnetization $M^{z}$ and
(b) isothermal susceptibility $\protect\chi _{T}^{zz},$at $T=0$ and for $%
n=8, $ $\protect\mu _{1}=....=\protect\mu _{8}=1,$ as functions of
the
uniform field for $J_{1}=1,$ $J_{2}=2,$ $J_{3}=1.5,$ $J_{4}=5/3,$ $%
J_{5}=2/3, $ $J_{6}=3/5,$ $J_{7}=2,$ $J_{8}=1$ (continuous line$)$ and $%
J_{1}=2,$ $J_{2}=1, $ $J_{3}=1.5,$ $J_{4}=5/3,$ $J_{5}=2/3,$ $J_{6}=3/5,$ $%
J_{7}=2,$ $J_{8}=1$ (dashed line).} \label{fig3}
\end{figure}

\pagebreak
\begin{figure}
\begin{center}
\includegraphics[width=14cm]{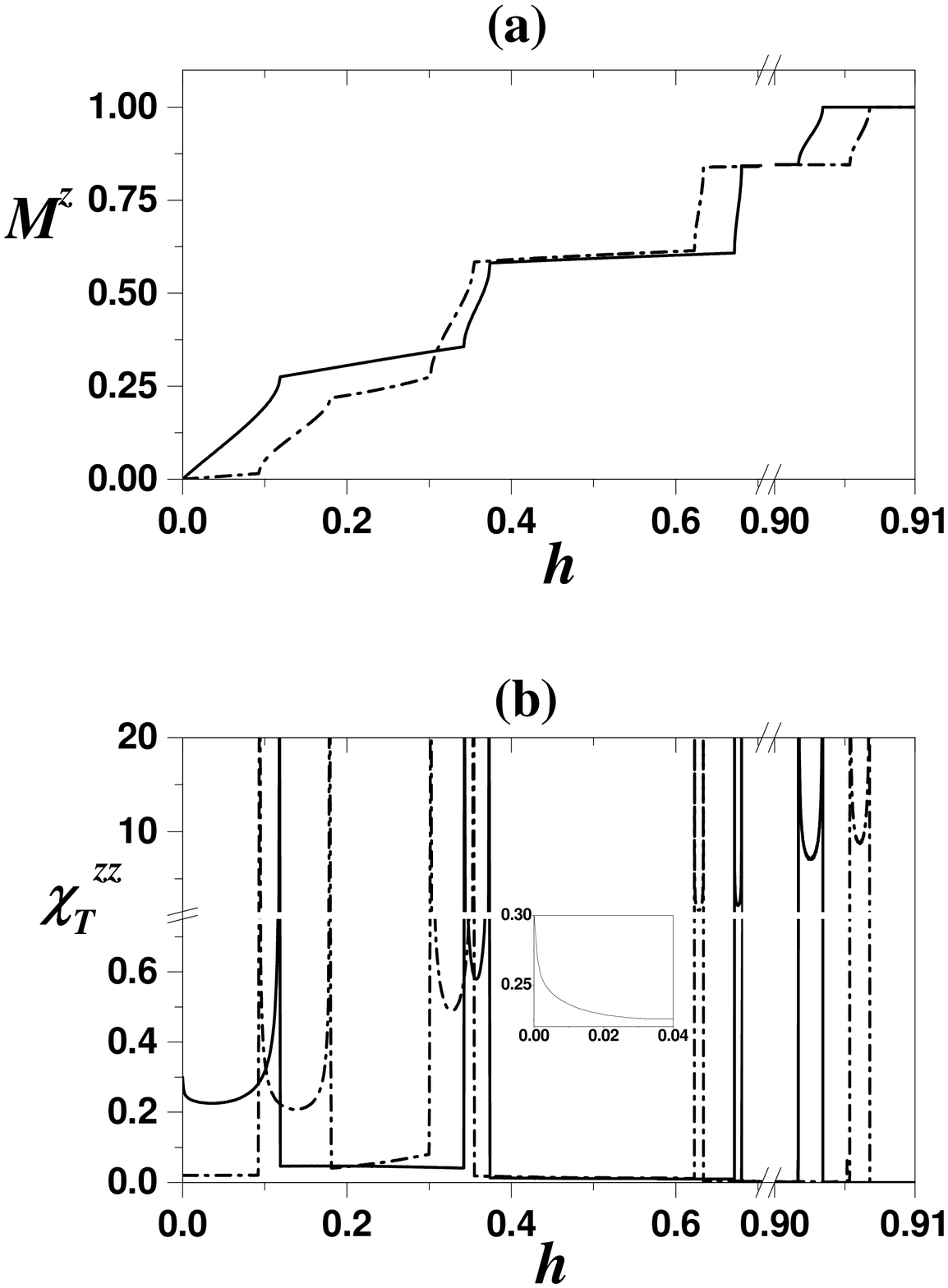}
\end{center}
\caption{(a) Magnetization $M^{z}$ and
(b) isothermal susceptibility $\protect\chi _{T}^{zz},$ at $T=0$ and $n=8,$ $%
\protect\mu _{1}=3,$ $\protect\mu _{2}=4,$ $\protect\mu _{3}=\protect\mu %
_{5}=\protect\mu _{7}=1.5,$ $\protect\mu _{4}=2.5,$ $\protect\mu _{6}=%
\protect\mu _{8}=1$ $,$ as functions of the uniform field for $J_{1}=1,$ $%
J_{2}=2,$ $J_{3}=1.5,$ $J_{4}=5/3,$ $J_{5}=2/3,$ $J_{6}=3/5,$ $J_{7}=2,$ $%
J_{8}=1$ (continuous line$)$ and $J_{1}=2,$ $J_{2}=1,$ $J_{3}=1.5,$ $%
J_{4}=5/3,$ $J_{5}=2/3,$ $J_{6}=3/5,$ $J_{7}=2,$ $J_{8}=1$ (dashed
line).} \label{fig4}
\end{figure}

\pagebreak
\begin{figure}
\begin{center}
\includegraphics[width=14cm]{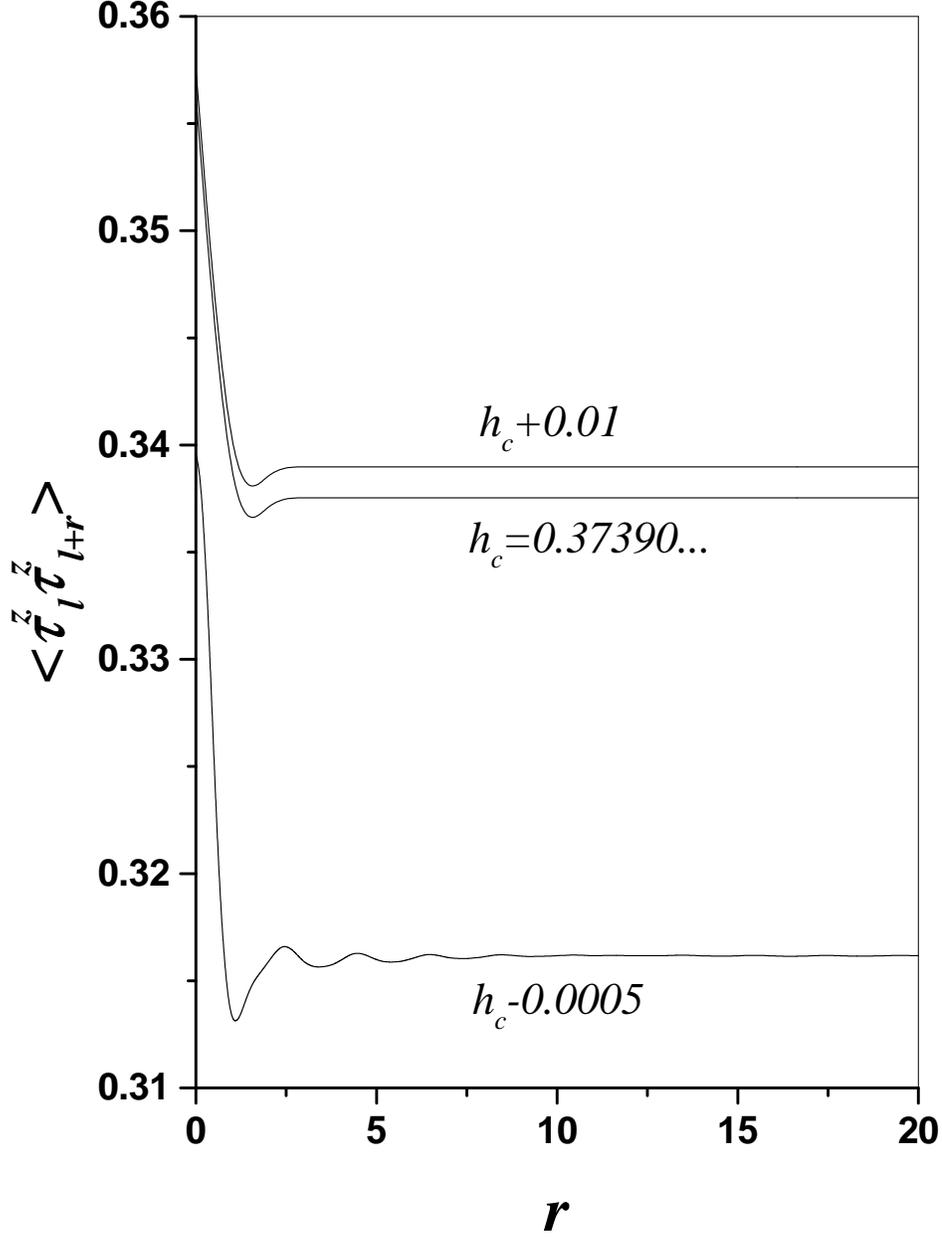}
\end{center}
\caption{Static correlation function $<\protect\tau
_{l}^{z}\protect\tau _{l+r}^{z}>$ as a function of $r
$ (distance between cells), at $T=0,$ for $n=8$, $J_{1}=1,$ $J_{2}=2,$ $%
J_{3}=1.5,$ $J_{4}=5/3,$ $J_{5}=2/3,$ $J_{6}=3/5,$ $J_{7}=2,$ $J_{8}=1$, $%
\protect\mu _{1}=3,$ $\protect\mu _{2}=4,$ $\protect\mu _{3}=\protect\mu %
_{5}=\protect\mu _{7}=1.5,$ $\protect\mu _{4}=2.5,$ $\protect\mu _{6}=%
\protect\mu _{8}=1$ for values of the field near and at the critical
field.} \label{fig5}
\end{figure}

\pagebreak
\begin{figure}
\begin{center}
\includegraphics[width=14cm]{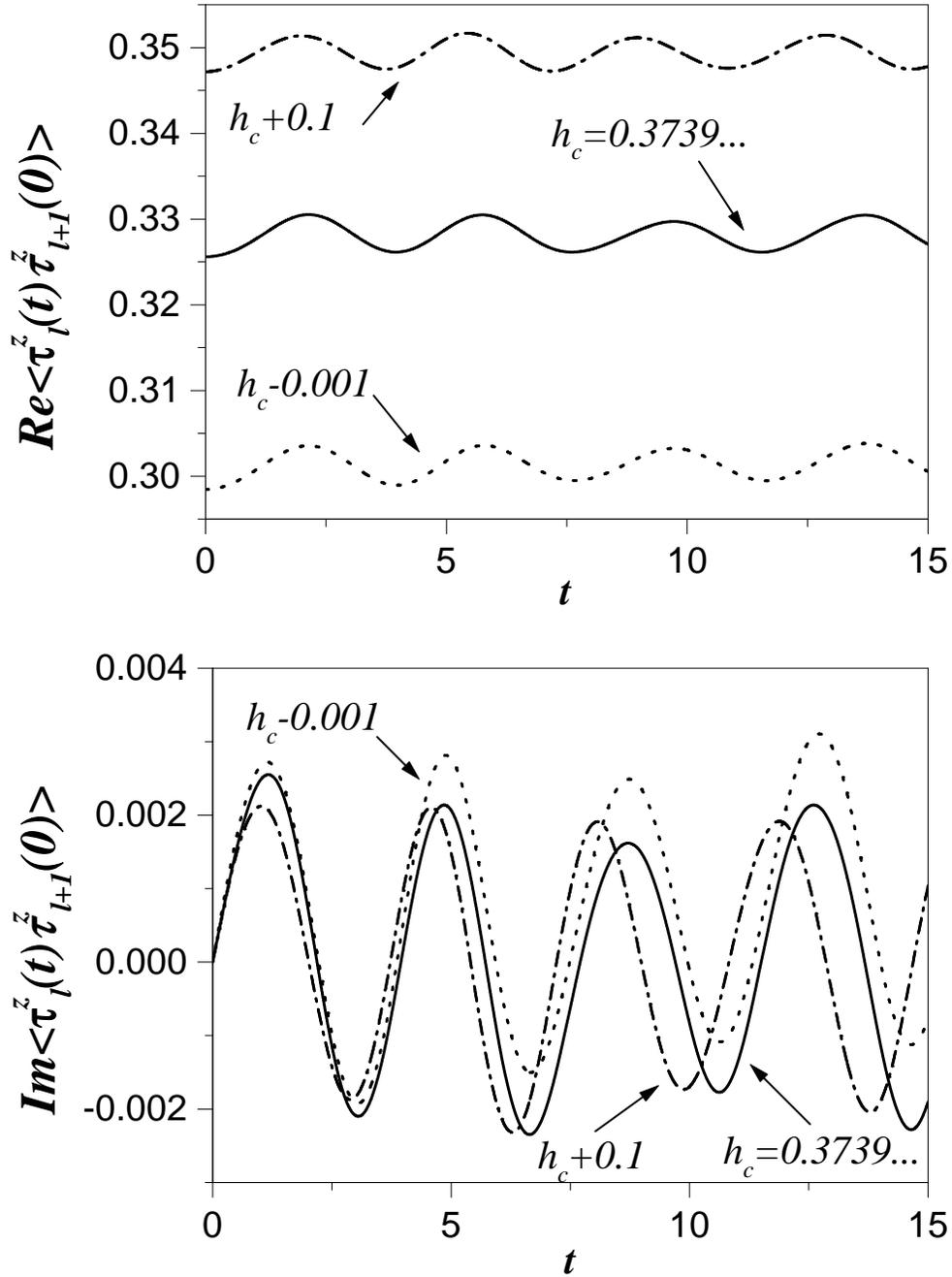}
\end{center}
\caption{The real and
imaginary parts of the correlation function $\left\langle \protect\tau %
_{l}^{z}(t)\protect\tau _{l+1}^{z}(0)\right\rangle ,$ as functions
of time,
at $T=0,$ for $n=8$, $J_{1}=1,$ $J_{2}=2,$ $J_{3}=1.5,$ $J_{4}=5/3,$ $%
J_{5}=2/3,$ $J_{6}=3/5,$ $J_{7}=2,$ $J_{8}=1$, $\protect\mu _{1}=3,$ $%
\protect\mu _{2}=4,$ $\protect\mu _{3}=\protect\mu _{5}=\protect\mu
_{7}=1.5, $ $\protect\mu _{4}=2.5,$ $\protect\mu _{6}=\protect\mu
_{8}=1,$ for values of the field near and at the critical field.}
\label{fig6}
\end{figure}

\pagebreak
\begin{figure}
\begin{center}
\includegraphics[width=14cm]{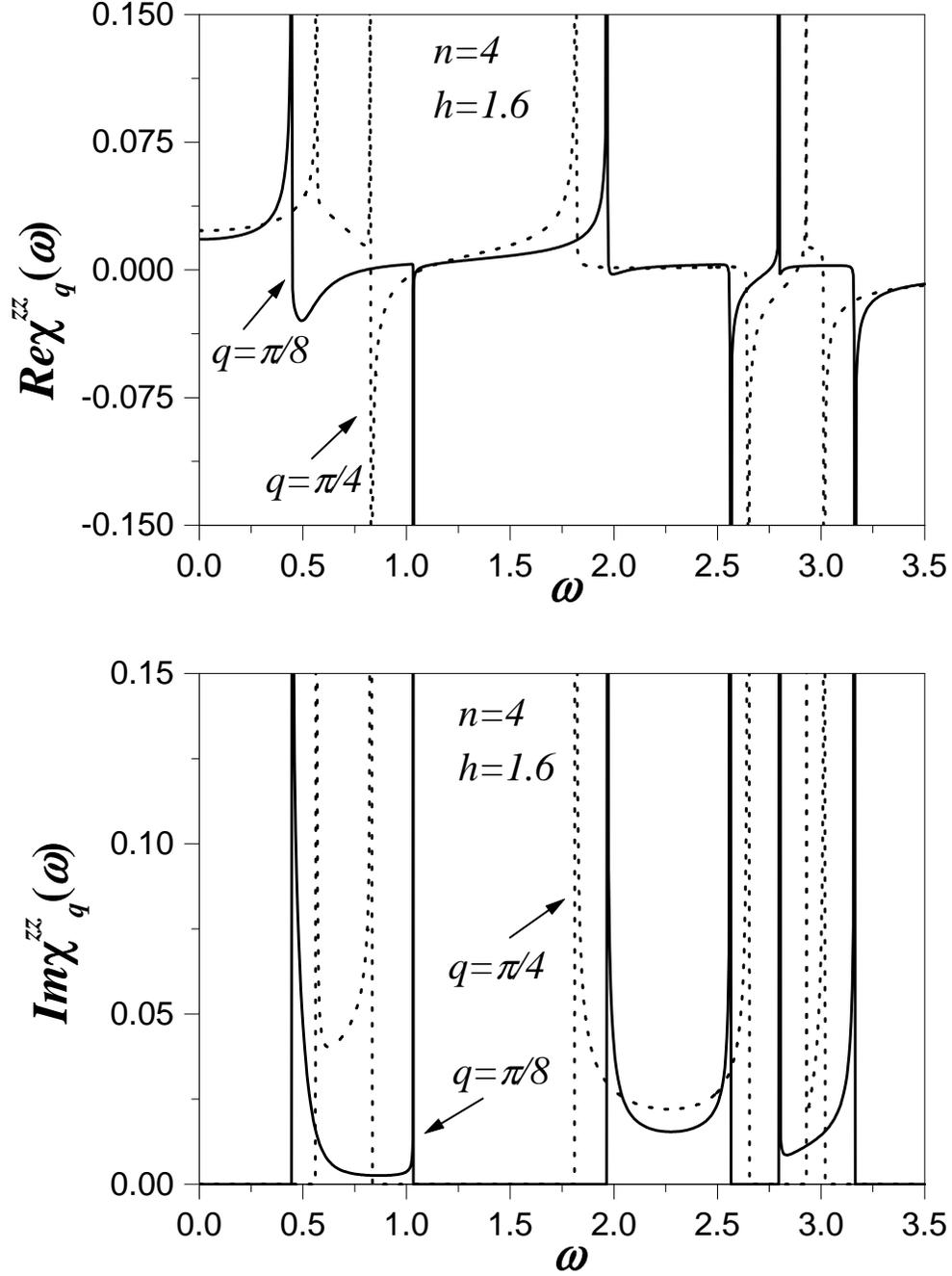}
\end{center}
\caption{The real and
imaginary parts of the dynamic susceptibility in the field direction, $%
\protect\chi _{q}^{zz}(\protect\omega ),$ at $T=0$ as a function of
frequency for $n=4$, $J_{1}=1,$ $J_{2}=2,$ $J_{3}=1.5,$ $J_{4}=5/3,$ $%
\protect\mu _{1}=\protect\mu _{3}=1$, $\protect\mu _{2}=\protect\mu
_{4}=0.5$ and $h=1,$for different values of $q$.} \label{fig7}
\end{figure}

\pagebreak
\begin{figure}
\begin{center}
\includegraphics[width=14cm]{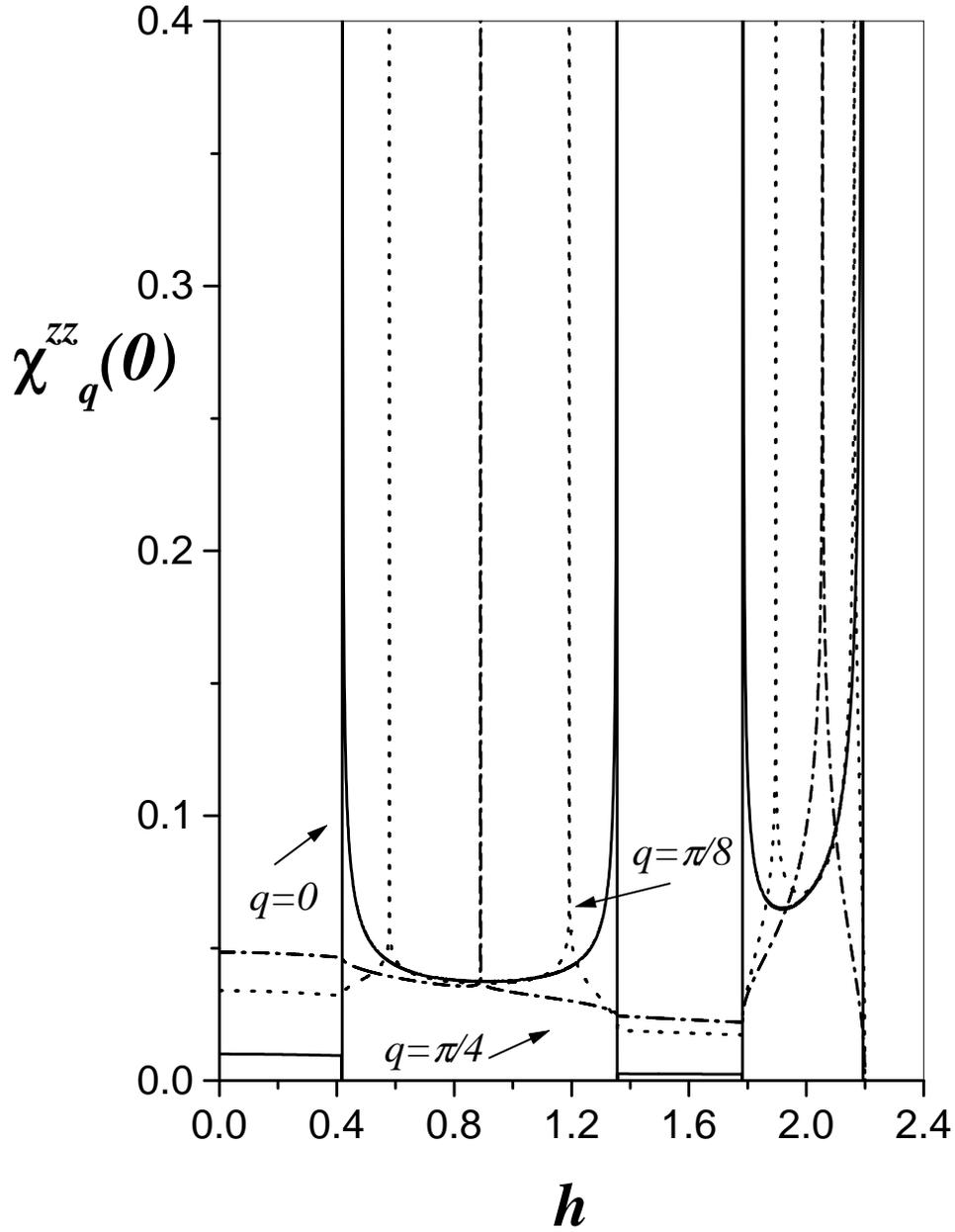}
\end{center}
\caption{Static susceptibility in the field direction, $\protect\chi
_{q}^{zz}(0)$, at $T=0,$
as a function of the field for $n=4$, $J_{1}=1,$ $J_{2}=2,$ $J_{3}=1.5,$ $%
J_{4}=5/3,$ $\protect\mu _{1}=1$, $\protect\mu _{2}=0.5,$ $\protect\mu %
_{3}=1,$ $\protect\mu _{4}=0.5$ and different values of $q$.}
\label{fig8}
\end{figure}

\pagebreak
\begin{figure}
\begin{center}
\includegraphics[width=14cm]{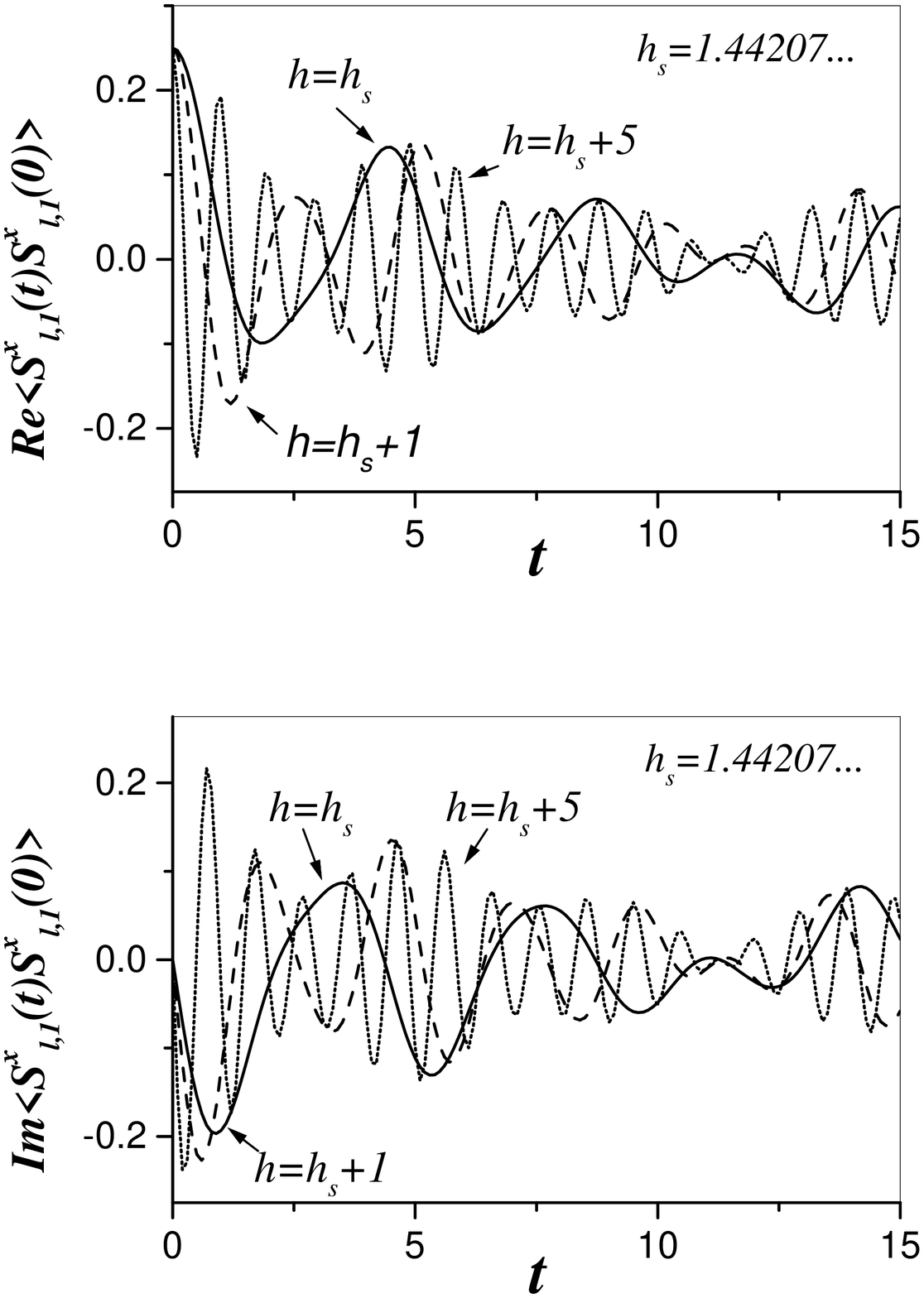}
\end{center}
\caption{The real and imaginary parts of the correlation function
$\left\langle S_{l,1}^{x}(t)S_{l,1}^{x}(0)\right\rangle ,$ as
functions of time, at $T=0,$
for $n=8$, $J_{1}=1,$ $J_{2}=2,$ $J_{3}=1.5,$ $J_{4}=5/3,$ $J_{5}=2/3,$ $%
J_{6}=3/5,$ $J_{7}=2,$ $J_{8}=1$, $\protect\mu _{1}=....=\protect\mu
_{8}=1,$ for values of the field at and at above the saturation
field $h_{s}$.} \label{fig9}
\end{figure}

\pagebreak
\begin{figure}
\begin{center}
\includegraphics[width=14cm]{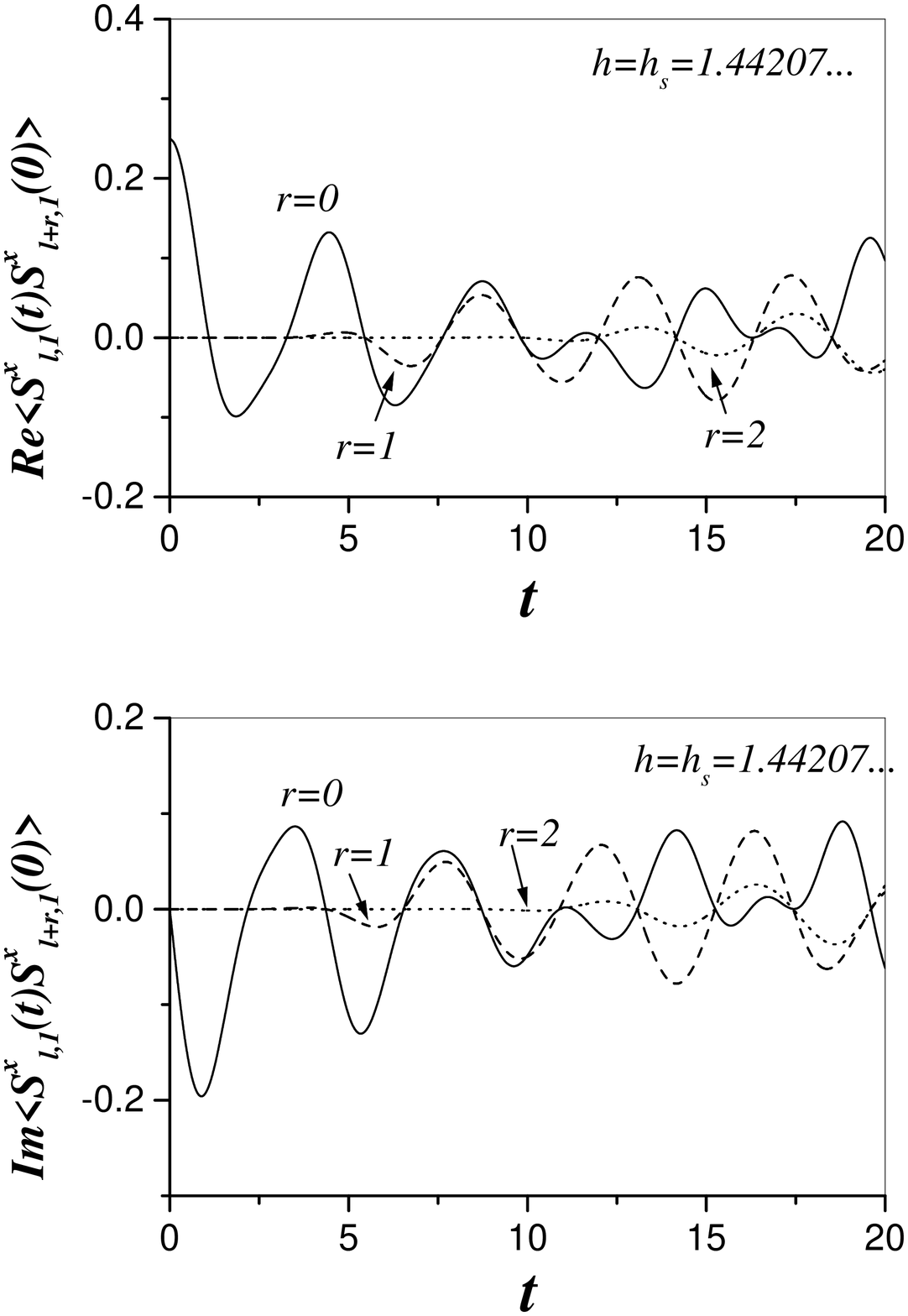}
\end{center}
\caption{The real and imaginary parts of the correlation function
$\left\langle
S_{l,1}^{x}(t)S_{l+r,1}^{x}(0)\right\rangle ,$ as functions of time, at $%
T=0, $ for $n=8$, $J_{1}=1,$ $J_{2}=2,$ $J_{3}=1.5,$ $J_{4}=5/3,$ $%
J_{5}=2/3, $ $J_{6}=3/5,$ $J_{7}=2,$ $J_{8}=1$, $\protect\mu _{1}=....=%
\protect\mu _{8}=1,$ at the saturation field $h_{s}$.} \label{fig10}
\end{figure}

\pagebreak
\begin{figure}
\begin{center}
\includegraphics[width=14cm]{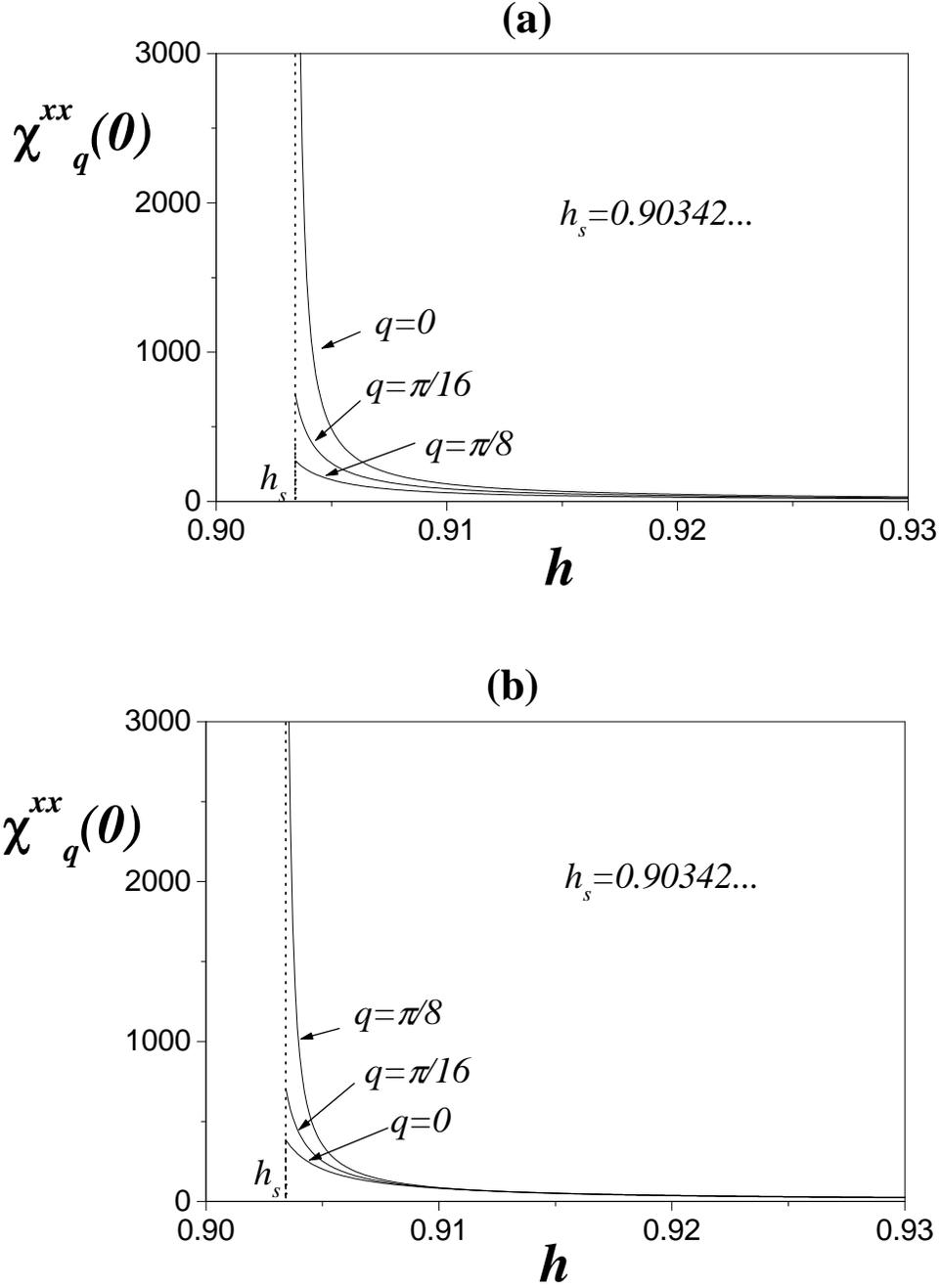}
\end{center}
\caption{Static
susceptibility $\protect\chi _{q}^{xx}(0)$ as a function of the field, at $%
T=0,$ for $n=8$, $\protect\mu _{1}=3,$ $\protect\mu _{2}=4,$ $\protect\mu %
_{3}=\protect\mu _{5}=\protect\mu _{7}=1.5,$ $\protect\mu _{4}=2.5,$ $%
\protect\mu _{6}=\protect\mu _{8}=1$ $,$ and different values of
$q$. (a)
Corresponds to $J_{1}=1,J_{2}=2,$ $J_{3}=1.5,$ $J_{4}=5/3,$ $J_{5}=2/3,$ $%
J_{6}=3/5,$ $J_{7}=2,$ $J_{8}=1$ and (b) $J_{1}=-1,J_{2}=2,$ $J_{3}=1.5,$ $%
J_{4}=5/3,$ $J_{5}=2/3,$ $J_{6}=3/5,$ $J_{7}=2,$ $J_{8}=1$.}
\label{fig11}
\end{figure}

\appendix\label{appendixA}

\section{Transfer matrix technique: excitation spectrum}

\label{matriz_transferencia}

The excitation spectrum of $H_{q},$
\begin{eqnarray}
H_{q} &=&-\sum_{m=1}^{n}\mu _{m}h\left( A_{q,m}^{\dagger }A_{q,m}-\frac{1}{2}%
\right) -  \notag \\
&&-\sum_{m=1}^{n-1}\frac{J_{m}}{2}\left[ A_{q,m}^{\dagger
}A_{q,m+1}+A_{q,m+1}^{\dagger }A_{q,m}\right] -  \notag \\
&&-\frac{J_{n}}{2}\left[ A_{q,n}^{\dagger }A_{q,1}\exp
(-iqd)+A_{q,1}^{\dagger }A_{q,n}\exp (iqd)\right] ,  \label{HqB}
\end{eqnarray}%
given in eq.(\ref{dispersion}), can also be obtained by using a
transfer
matrix method. By introducing the equation of motion for the operators $%
A_{q,m}$
\begin{equation}
\dot{A}_{q,m}=\frac{1}{i}[A_{q,m},\sum_{q^{\prime }}H_{q^{\prime
}}], \label{motion equation}
\end{equation}%
and assuming the time-evolution given by
\begin{equation}
A_{q,m}(t)=A_{q,m}\exp (-i\omega t),
\end{equation}%
we obtain the system of equations

\begin{equation}
\begin{split}
(\omega +\mu _{1}h)A_{q,1}& =-\frac{J_{1}}{2}A_{q,2}-\frac{J_{n}\exp (iqd)}{2%
}A_{q,n}, \\
(\omega +\mu _{2}h)A_{q,2}&
=-\frac{J_{2}}{2}A_{q,3}-\frac{J_{1}}{2}A_{q,1},
\\
& \vdots \\
(\omega +\mu _{m-1}h)A_{q,m-1}& =-\frac{J_{m-1}}{2}A_{q,m}-\frac{J_{m-2}}{2}%
A_{q,m-2}, \\
(\omega +\mu _{m}h)A_{q,m}& =-\frac{J_{m}}{2}A_{q,m+1}-\frac{J_{m-1}}{2}%
A_{q,m-1}, \\
(\omega +\mu _{m+1}h)A_{q,m+1}& =-\frac{J_{m+1}}{2}A_{q,m+2}-\frac{J_{m}}{2}%
A_{q,m}, \\
& \vdots \\
(\omega +\mu _{n-1}h)A_{q,n-1}& =-\frac{J_{n-1}}{2}A_{q,n}-\frac{J_{n-2}}{2}%
A_{q,n-2}, \\
(\omega +\mu _{n}h)A_{q,n}& =-\frac{J_{n}\exp (-iqd)}{2}A_{q,1}-\frac{J_{n-1}%
}{2}A_{q,n-1},
\end{split}%
\end{equation}%
which is equivalent to the eigenvalue equation shown in eq.(\ref%
{eigenequation}). From this set of equations we can write the matrix
equation

\begin{equation}
\begin{pmatrix}
A_{q,m+1} \\
A_{q,m}%
\end{pmatrix}%
=\mathbb{T}_{m}(\omega ,h)%
\begin{pmatrix}
A_{q,m} \\
A_{q,m-1}%
\end{pmatrix}%
,\text{ for }m=2,3,...,n-1,  \label{matricial equationB}
\end{equation}%
where

\begin{equation}
\mathbb{T}_{m}(\omega ,h)\equiv
\begin{pmatrix}
-2(\frac{\omega +\mu _{m}h}{J_{m}}) & -\frac{J_{m-1}}{J_{m}} \\
1 & 0%
\end{pmatrix}%
,\text{ }  \label{tmB}
\end{equation}%
and
\begin{eqnarray}
\begin{pmatrix}
A_{q,2} \\
A_{q,1}%
\end{pmatrix}
&=&\mathbb{\widetilde{T}}_{1}(\omega ,h)%
\begin{pmatrix}
A_{q,1} \\
A_{q,n}%
\end{pmatrix}%
\text{ for }m=1,  \label{m=1} \\
\begin{pmatrix}
A_{q,1} \\
A_{q,n}%
\end{pmatrix}
&=&\mathbb{\widetilde{T}}_{n}(\omega ,h)%
\begin{pmatrix}
A_{q,n} \\
A_{q,n-1}%
\end{pmatrix}%
,\text{ for }m=n  \label{m=n}
\end{eqnarray}%
with $\mathbb{\widetilde{T}}_{1}$ and $\mathbb{\widetilde{T}}_{n}$
given by
\begin{eqnarray}
\mathbb{\widetilde{T}}_{1}(\omega ,h) &\equiv &%
\begin{pmatrix}
-2(\frac{\omega +\mu _{1}h}{J_{1}}) & -\frac{J_{n}}{J_{1}}\exp (idq) \\
1 & 0%
\end{pmatrix}%
,  \label{t1tilde} \\
\mathbb{\widetilde{T}}_{n}(\omega ,h) &\equiv &\exp (idq)%
\begin{pmatrix}
-2(\frac{\omega +\mu _{n}h}{J_{n}}) & -\frac{J_{n-1}}{J_{n}} \\
\exp (-idq) & 0%
\end{pmatrix}%
,  \label{t2tilde}
\end{eqnarray}%
which satisfy the result,
\begin{equation}
\mathbb{\widetilde{T}}_{1}(\omega
,h)\mathbb{\widetilde{T}}_{n}(\omega ,h)=\exp
(inq)\mathbb{T}_{1}(\omega ,h)\mathbb{T}_{n}(\omega ,h),
\label{t1tilde*t2tilde}
\end{equation}%
with $\mathbb{T}_{1}$ and $\mathbb{T}_{n}$ given by
\begin{eqnarray}
\mathbb{T}_{1}(\omega ,h) &\equiv &%
\begin{pmatrix}
-2(\frac{\omega +\mu _{1}h}{J_{1}}) & -\frac{J_{n}}{J_{1}} \\
1 & 0%
\end{pmatrix}%
,  \label{t1} \\
\mathbb{T}_{n}(\omega ,h) &\equiv &%
\begin{pmatrix}
-2(\frac{\omega +\mu _{n}h}{J_{n}}) & -\frac{J_{n-1}}{J_{n}} \\
1 & 0%
\end{pmatrix}%
,  \label{t2}
\end{eqnarray}%
From eqs.(\ref{matricial equationB}-\ref{t2}) we can write
\begin{equation}
\mathbb{T}_{cell}(\omega ,h)%
\begin{pmatrix}
A_{q,2} \\
A_{q,1}%
\end{pmatrix}%
=\exp (-iqd)%
\begin{pmatrix}
A_{q,2} \\
A_{q,1}%
\end{pmatrix}%
\equiv \exp (-iqd)\Psi _{q,1,2},  \label{transfer}
\end{equation}%
where
\begin{equation}
\mathbb{T}_{cell}(\omega ,h)\equiv \mathbb{T}_{1}(\omega ,h)\mathbb{T}%
_{n}(\omega ,h)\mathbb{T}_{n-1}(\omega ,h)\mathbb{T}_{n-2}(\omega ,h)\hdots%
\mathbb{T}_{3}(\omega ,h)\mathbb{T}_{2}(\omega ,h).  \label{tcell}
\end{equation}%
Eq.(\ref{transfer}) shows explicitly that $\exp (-iqn)$ is an eigenvalue of $%
\mathbb{T}_{cell}(\omega ,h)$ corresponding to the eigenvector $\Psi
_{q,1,2}.$ Since $\mathbb{T}_{cell}(\omega ,h)$ does not depend on
$q,$we can obtain immediately the second eigenvector of
$\mathbb{T}_{cell}(\omega ,h)$, and \ the respective eigenvalue,
from this equation by introducing the transformation $q\rightarrow
-q,$ which gives$\Psi _{-q,1,2}$ and $\exp
(-iqd)$ respectively. These results are consistent with the fact that $det[%
\mathbb{T}_{cell}(\omega ,h)]=1,$ and we can write finally the
equation
\begin{equation}
trace[\mathbb{T}_{cell}(\omega ,h)]=2\cos (dq),  \label{dispersionB}
\end{equation}%
whose solution will give the excitation spectrum.

In the absence of the external field, for $n$ odd, we can show that
the previous equation can be written in the form

\begin{equation}
f(\omega )\times \omega =2\cos (dq),
\end{equation}%
where $f(\omega )$ is a polynomial function of degree $n-1.$ This
means that $q=\pi /2d$ is a zero-energy mode irrespective of the
values of $J^{\prime }s.$

On the other hand, for $n$ even and zero external field, eq.(\ref%
{dispersionB}) can be written as

\begin{equation}
g(\omega )\times \omega ^{2}+(-1)^{n/2}\left[ \frac{J_{2}J_{4}...J_{n}}{%
J_{1}J_{3}...J_{n-1}}+\frac{J_{1}J_{3}...J_{n-1}}{J_{2}J_{4}...J_{n}}\right]
=2\cos (dq),  \label{gapcondition}
\end{equation}%
where $g(\omega )$ is a polynomial of degree $n-2.$ From this result
we can
conclude that $q=0$ is a zero-energy mode provided the condition $%
J_{1}J_{3}...J_{n-1}=J_{2}J_{4}...J_{n}$ is satisfied. Therefore,
under this condition, there is no energy gap between the ground
state and the first excited state in the absence of external field.

\section{Dynamic correlation $\left\langle S_{l,m}^{x}(t)S_{l+r,m^{\prime
}}^{x}(0)\right\rangle $}

\label{appendixB} 

The dynamic correlation function $\left\langle
S_{l,m}^{x}(t)S_{l+r,m^{\prime }}^{x}(0)\right\rangle ,$ in terms of
the hamiltonians $H^{\pm }$, in the thermodynamic limit, can be
written in the
form \cite{siskens:1974,capel:1977,goncalvestese:1977}%
\begin{equation}
\left\langle S_{l,m}^{x}(t)S_{l+r,m^{\prime }}^{x}(0)\right\rangle =\frac{Tr%
\left[ \exp (-\beta H^{-})\exp (iH^{-}t)S_{l,m}^{x}\exp
(-iH^{+}t)S_{l+r,m^{\prime }}^{x}\right] }{Tr\left[ \exp (-\beta H^{-})%
\right] },
\end{equation}%
which is very difficult to calculate, since $H^{-}$ and $H^{+}$ do
not commute. However, at $T=0$, for external field greater than the
saturation field $(h\geqslant h_{s}),$ $H^{+}$ and $H^{-}$ have
identical ground state,
namely,%
\begin{equation}
\left\vert \Phi _{0}\right\rangle =\prod_{l,m}\otimes \left\vert
n_{lm}\right\rangle  \label{groundstate}
\end{equation}%
where $n_{lm}=1,$ $\forall $ $l,m,$ and
\begin{equation}
H^{-}\left\vert \Phi _{0}\right\rangle =H^{+}\left\vert \Phi
_{0}\right\rangle =E_{0}\left\vert \Phi _{0}\right\rangle .
\end{equation}%
Then, in this limit, the dynamic correlation $\left\langle
S_{1,m}^{x}(t)S_{1+r,m^{\prime }}^{x}(0)\right\rangle $ is given by
\begin{equation}
\langle S_{1,m}^{x}(t)S_{1+r,m^{\prime }}^{x}(0)\rangle
=\left\langle \Phi _{0}\right\vert \exp (iH^{+}t)S_{1,m}^{x}\exp
(-iH^{-}t)S_{1+r,m^{\prime }}^{x}\left\vert \Phi _{0}\right\rangle .
\label{corrxx1}
\end{equation}%
From the eqs.(\ref{ladder}) and (\ref{Jordan-Wigner}) we can express $%
S_{l,m}^{x}$ in terms of fermion operators in the form%
\begin{equation}
S_{l,m}^{x}=\frac{1}{2}\exp \left\{ i\pi \left[ \sum_{j=1}^{r}%
\sum_{k=1}^{n}c_{j,k}^{\dag }c_{j,k}+\sum_{k^{\prime }=1}^{m^{\prime
}-1}c_{1+r,k^{\prime }}^{\dag }c_{1+r,k^{\prime }}\right] \right\}
\left( c_{1+r,m^{\prime }}^{\dag }+c_{1+r,m^{\prime }}\right) ,
\end{equation}%
and substituting this result in eq.(\ref{corrxx1}) we obtain

\begin{eqnarray}
\left\langle S_{1,m}^{x}(t)S_{1+r,m^{\prime }}^{x}(0)\right\rangle &=&\frac{%
(-1)^{nr+m^{\prime }-1}}{4}\exp (iE_{0}t)(-1)^{m-1}\left\langle \Phi
_{0}\right\vert \left( c_{1,m}^{\dag }+c_{1,m}\right) \times  \notag \\
&&\times \exp (-iH^{-}t)\left( c_{1+r,m^{\prime }}^{\dag
}+c_{1+r,m^{\prime }}\right) \left\vert \Phi _{0}\right\rangle ,
\end{eqnarray}%
which can be written as%
\begin{eqnarray}
\left\langle S_{1,m}^{x}(t)S_{1+r,m^{\prime }}^{x}(0)\right\rangle &=&\frac{%
(-1)^{nr+m^{\prime }-m}}{4}\left\langle \Phi _{0}\right\vert \exp
(iH^{-}t)[c_{1,m}^{\dag }+c_{1,m}]\exp (-iH^{-}t)\times  \notag \\
&&\times \left[ c_{1+r,m^{\prime }}^{\dag }+c_{1+r,m^{\prime
}}\right] \left\vert \Phi _{0}\right\rangle .
\end{eqnarray}%
Finally, by using eq.(\ref{contractions}), we obtain
\begin{equation}
\left\langle S_{1,m}^{x}(t)S_{1+r,m^{\prime }}^{x}(0)\right\rangle =\frac{%
(-1)^{nr+m^{\prime }-m}}{4N}\sum_{q,k}\exp (-iqdr)u_{q,km}^{\ast
}u_{q,km^{\prime }}\exp (i\varepsilon _{q,k}t),
\label{correlationxx}
\end{equation}%
which reduces to the known result for the homogeneous chain
\cite{cruz:1981}.

\bigskip From the previous equation we can conclude immediately that the
static correlation $\left\langle S_{1,m}^{x}S_{1+r,m^{\prime
}}^{x}\right\rangle $ is equal to $\delta _{q,0}/4.$


\begin{thebibliography}{99}
\bibitem{dagotto:1996} E. Dagotto and T. M. Rice, Surprises on the way from
one- to two-dimensional quantum magnets: the ladder materials, \textit{%
Science} \textbf{271} (1996) 618-623.

\bibitem{nguyen:1996} T. N. Nguyen, P. A. Lee and H. C. Loye, Design of a
random quantum spin chain paramagnet:
$Sr_{3}CuPt_{0.5}Ir_{0.5}O_{6}$, \textit{Science} \textbf{271}
(1996) 489-491.

\bibitem{gambardella:2002} P. Gambardella, A. Dallmeyer, K. Maiti, M. C.
Malagoli,W. Eberhardt, K. Kern and C. Carbone, Ferromagnetism in
one-dimensional monoatomic metal chains, \textit{Nature}
\textbf{416} (2002) 301-303.

\bibitem{mukherjeea:2004} C.J. Mukherjeea, R. Coldeaa, D.A. Tennanta, M.
Kozac, M. Enderlec, K. Habichtd, P. Smeibidld, and Z. Tylczynskie,
Field-induced quantum phase transition in the quasi 1D XY-like
antiferromagnet $Cs_{2}CoCl_{4},$ \textit{J. Magn. Magn. Matt}, \textbf{%
272--276} (2004) 920--921.

\bibitem{sachdev:2000} S. Sachdev, \textit{Quantum phase transitions}
(Cambridge Univ. Press, 2000).

\bibitem{okamoto:2002} K. Okamoto, Level spectroscopy: physical meaning and
application to the magnetization plateau problems, \textit{Prog.
Theoret. Phys. Supp}.\textbf{\ 145} (2002) 113-118.

\bibitem{bostrem:2003} I. G. Bostrem, A. S. Boyarchenkov, A. A. Konovalov,
A. S. Ovchinnikov and V. E. Sinitsyn, \textit{J. Exp. and Theoret.
Phys.} \textbf{97} (2003) 615--623.

\bibitem{kramp:2000} S. Kramp, M. Loewenhaupt and M. Rotter, The spin wave
dispersion of $NdCu2$ in strong magnetic fields,\textit{\ Physica
B,} \textbf{276-278} (2000) 628-629.

\bibitem{coleman:2001} P. Coleman, Magnetic spins that last for ever,
\textit{Nature} \textbf{413 }(2001) 788-789.

\bibitem{matsumoto:2004} M. Matsumoto, B. Normand, T. M. Rice and M.
Sigrist, Field- and pressure-induced magnetic quantum phase transitions in $%
TlCuCl_{3},$\textit{Phys. Rev. B} \ \textbf{69} (2004) 054423.

\bibitem{lieb:1961} H. E. Lieb, T. Schultz and D. C. Mattis, Two soluble
models of antiferromagnetic chain, \textit{Ann. Phys}. \textbf{16
}(1961) 407-466.

\bibitem{derzhko:2004} O. Derzhko, J. Richter, T. Krokhmalskii and O.
Zaburannyi, Regularly alternating spin-1/2 anisotropic XY chains:
The
ground-state and thermodynamic properties, \textit{Phys.Rev.E} \textbf{69 }%
(2004)\textbf{\ }066112.

\bibitem{derzhko:2000} O. Derzhko, J. Richter and O. Zaburannyi,
Thermodynamic properties of the periodic nonuniform spin-1/2
isotropic XY chains in a transverse field, \textit{Physica A}
\textbf{282} (2000) 495-524.

\bibitem{kontorovich:1968} V. M. Kontorovich and V. M. Tsukernik, Magnetic
properties of a spin array with two sublattices, \textit{Sov. Phys.
JETP} \textbf{26} (1968) 687-691.

\bibitem{perk:1980} J. H. H. Perk and H. W. Capel, Time- and
frequency-dependent correlation functions for the homogeneous and
alternating isotropic XY-models, \textit{Physica} \textbf{100A}
(1980) 1-23.

\bibitem{derzkho:2000dyn} O. Derzhko, T. Krokhmalskii and J. Stolze,
Dynamics of the spin-1/2 isotropic XY chain in a transverse field, \textit{%
J. Phys. A: Math. Gen}. \textbf{33} (2000) 3063-3080.

\bibitem{delima:2002} J. P. de Lima and L. L. Gon\c{c}alves, The
longitudinal dynamic correlation and dynamic susceptibility of the
isotropic XY-model on the 1D alternating superlattice,
\textit{Physica A }\textbf{311} (2002) 458-474\textit{.}

\bibitem{delima:1999} J. P. de Lima and L. L. Gon\c{c}alves, The XY model on
the one-dimensional superlattice: static properties, \textit{J.
Magn. Magn. Mater}. \textbf{206} (1999) 135-148.

\bibitem{goncalves:1995} L. L. Gon\c{c}alves and J. P. de Lima, The XY-model
on the one-dimensional superlattice, \textit{J. Magn. Magn. Mater}. \textbf{%
140-144} (1995) 1606-1608.

\bibitem{siskens:1974} Th. J. Siskens and P. Mazur, Time-correlation
functions in the a-ciclic XY model, \textit{Physica} \textbf{71
}(1974) 560-578.

\bibitem{capel:1977} H. W. Capel and J. H. H. Perk, Autocorrelation function
of x-component of the magnetization in the one-dimensional XY-model, \textit{%
Physica A} \textbf{87} (1977) 211-242.

\bibitem{goncalvestese:1977} L. L. Goncalves, \textit{Theory of properties
of some one-dimensional systems} (D.\ Phil. Thesis, University of
Oxford, 1977).

\bibitem{barbosafilho:2001} F. F. Barbosa Filho, J. P. de Lima and L. L. Gon%
\c{c}alves, The anisotropic XY model on the 1D alternating
superlattice, \textit{J. Magn. Magn. Mater}. \textbf{226-230} (2001)
638-640.

\bibitem{korn:1961} G. A. Korn and \ T. M. Korn, \textit{Mathematical
handbook for scientists and engineers }(Mc Graw-Hill, New York,
1961).

\bibitem{pfeuty:1979} P. Pfeuty, An exact result for the 1D random Ising
model in a transverse field, \textit{Phys. Lett}.\textit{\ A}
\textbf{72} (1979) 245-246.

\bibitem{jullien:1978} R. Jullien and J. N. Fields, Equivalence between \ a
spin 1/2 dimerized XY chain and two independent Ising chains in
transverse fields, \textit{Phys. Lett.} \textit{A} \textbf{69
}(1978) 214-216.

\bibitem{dfisher:1994} D. S. Fisher, Random antiferromagnetics quantum spins
chains, \textit{Phys. Rev. B}\ \textbf{50} (1994) 3799-3821.

\bibitem{oshikawa:1997} M. Oshikawa, M. Yamanaka and I. Affleck,
Magnetization plateaus in spin chains: "Haldane gap" for
half-integer spins, \textit{Phys. Rev. Lett.} \textbf{78} (1997)
1984--1987.

\bibitem{niemeijer:1967} Th. Niemeijer, Some exact calculations on a chain
of spin $1/2$, \textit{Physica} \textbf{37 }(1967) 377-419.

\bibitem{parry:1973} W. E. Parry, \textit{The many-body problem} (Oxford
University Press, Oxford, 1973).

\bibitem{zubarev:1960} D. N. Zubarev, Double-time Green functions in
statistical physics, \textit{Sov. Phys. Usp.} \textbf{3 }(1960)
320-345.

\bibitem{katsura:1970} S. Katsura and T. Horiguchi, Dynamic properties of
the isotropic XY model, \textit{Physica} \textbf{46} (1970) 67-86.

\bibitem{jullien:1979} R. Jullien and P. Pfeuty, Zero-temperature
renormalization-group method for quantum systems. II. Isotropic X-Y
model in a transverse field in one dimension, \textit{Phys. Rev B}
\textbf{19} (1979) 4646-4652.

\bibitem{goncalves:1986} L. L. Gon\c{c}alves, Transverse susceptibility of
1D isotropic XY-model at zero temperature, \textit{Rev. Bras. Fis.} \textbf{%
16} (1986) 491-494.

\bibitem{horiguchi:1975} T. Horiguchi and T. Morita, Isothermal and
frequency dependent perpendicular susceptibilities of the Ising
model on the Cayley tree, \textit{Can. J. Phys}.\textbf{\ 53} (1975)
2375-2386.

\bibitem{stanley:1971} H. E. Stanley, \textit{Phase transitions and critical
phenomena} (Oxford University Press, Oxford, 1971).

\bibitem{delima:1997} J. P. de Lima and L. L. Gon\c{c}alves, Surface tension
of the induced magnetization in the 1D isotropic XY-model,
\textit{Mod. Phys. Lett}. B \textbf{11} (1997) 9-15.

\bibitem{cruz:1981} H. B. Cruz and L. L. Gon\c{c}alves, Time-dependent
correlation of the one-dimensional isotropic XY model, \textit{J.
Phys. C} \textbf{14} (1981) 2785-2791
\end{thebibliography}
\end{document}